\documentclass[journal]{IEEEtran}

\usepackage{amsmath}
\usepackage{amssymb}
\usepackage{amsfonts}
\usepackage{epsfig}
\usepackage{cite}
\usepackage{calc}
\usepackage{color}

\usepackage{hyperref}
\usepackage{times}
\usepackage{subfigure}
\DeclareGraphicsExtensions{.eps}
\usepackage[all,2cell]{xy} \UseAllTwocells \SilentMatrices

\newcommand{\R}{{\mathbb R}}

\newcommand{\Z}{{\mathbb Z}}
\newcommand{\N}{{\mathbb N}}

\newcommand{\G}{{\mathcal G}}
\newcommand{\T}{{\mathbb T}} 
\newcommand{\No}{\Z^+}

\newcommand{\ep}{\epsilon}

\newcommand{\ba}{\begin{array}}
\newcommand{\ea}{\end{array}}

\newcommand{\limrho}{\lim_{\rho\rightarrow\infty}}

\newcommand{\limN}{\lim_{N\rightarrow\infty}}
\newcommand{\SNR}{\text{SNR}}

\newcommand{\eqg}{\overset{g}{=}}
\newcommand{\geg}{\overset{g}{\ge}}
\newcommand{\leg}{\overset{g}{\le}}
\newcommand{\Pdv}{P_\text{delay}}
\newcommand{\Pt}{P_\text{tot}}
\newcommand{\Pch}{P_\text{ch}}
\newcommand{\dch}{d_\text{ch}}
\newcommand{\Icpe}{I} 

\newcommand{\Ng}{{(N)}}
\newcommand{\rmax}{r_\text{max}}
\newcommand{\ro}{r^*_{ir}}

\newtheorem{defn}{Definition}
\newtheorem{thm}{Theorem}

\newtheorem{lem}{Lemma}
\newtheorem{prop}[lem]{Proposition}

\newtheorem{note}{Remark}

\newtheorem{ex}{{\em Example}}

\newtheorem{approximation}{Approximation}

\newcommand{\no}{\nonumber}
\newcommand{\beq}{\begin{equation}}
\newcommand{\eeq}{\end{equation}}
\newcommand{\bmu}{\begin{multline}}
\newcommand{\emu}{\end{multline}}
\newcommand{\bmun}{\begin{multline*}}
\newcommand{\emun}{\end{multline*}}

\newcommand{\beqn}{\[}
\newcommand{\eeqn}{\]}
\newcommand{\bea}{\begin{eqnarray}}
\newcommand{\eea}{\end{eqnarray}}
\newcommand{\bean}{\begin{eqnarray*}}
\newcommand{\eean}{\end{eqnarray*}}

\newcommand{\bit}{\begin{itemize}}
\newcommand{\eit}{\end{itemize}}
\newcommand{\ben}{\begin{enumerate}}
\newcommand{\een}{\end{enumerate}}

\newcommand{\reqn}[1]{(\ref{#1})}

\newcommand{\ska}{\color{black}}
\newcommand{\skb}{\color{black}}
\newcommand{\ignore}[1]{}

\title{High-SNR Analysis of Outage-Limited Communications with Bursty and Delay-Limited Information}

\author{Somsak Kittipiyakul, Petros Elia, and Tara Javidi
\thanks{This work was supported in part by the Center for Wireless Communications, UCSD and UC Discovery Grant No. Com04-10176, ARO-MURI Grant No. W911NF-04-1-0224, NSF CAREER Award No. CNS-0533035, STREP project No. IST-026905 (MASCOT).
and AFOSR Grant No. FA9550-05-01-0430. The material in this paper was presented in part at the 5th International Symposium on Modeling and Optimization in Mobile, Ad Hoc, and Wireless Networks, Limassol, Cyprus, April 2007, the IEEE International Symposium on Information Theory, Nice, France, June 2007, and the 45th Annual Allerton Conference on Communication, Control, and Computing, Illinois, USA, September 2007.}
\thanks{S. Kittipiyakul and T. Javidi are with the Electrical and Computer Engineering Department,
University of California - San Diego, La Jolla, CA, 92093 (e-mail: \{skittipi,tjavidi\}@ucsd.edu).}
\thanks{P. Elia was with the Electrical and Computer Engineering Department,
University of California - San Diego, La Jolla, CA, 92093. He is now with the Mobile Communications System Department, EURECOM, Sophia Antipolis cedex, France (e-mail: elia@eurecom.fr).}
}

\begin{document}

\thispagestyle{empty}
 \maketitle


\begin{abstract}
This work analyzes the high-SNR asymptotic error performance of outage-limited communications with fading, where the number of bits that arrive at the transmitter during any time slot is random but the delivery of bits at the receiver must adhere to a strict delay limitation.  Specifically, bit errors are caused by erroneous decoding at the receiver or violation of the strict delay constraint. Under certain scaling of the statistics of the bit-arrival process with SNR, this paper shows that the optimal decay behavior of the asymptotic total probability of bit error depends on how fast the burstiness of the source scales down with SNR. If the source burstiness scales down too slowly, the total probability of error is asymptotically dominated by delay-violation events. On the other hand, if the source burstiness scales down too quickly, the total probability of error is asymptotically dominated by channel-error events. However, at the proper scaling, where the burstiness scales linearly with $\frac{1}{\sqrt{\log\SNR}}$ and at the optimal coding duration and transmission rate, the occurrences of channel errors and delay-violation errors are asymptotically balanced.
In this latter case, the optimal exponent of the total probability of error reveals a tradeoff that addresses the question of how much of the allowable time and rate should be used for gaining reliability over the channel and how much for accommodating the burstiness with delay constraints.
\end{abstract}

\section{Introduction}
\begin{figure*}[t]
\begin{center}
\includegraphics[width=.8\linewidth]{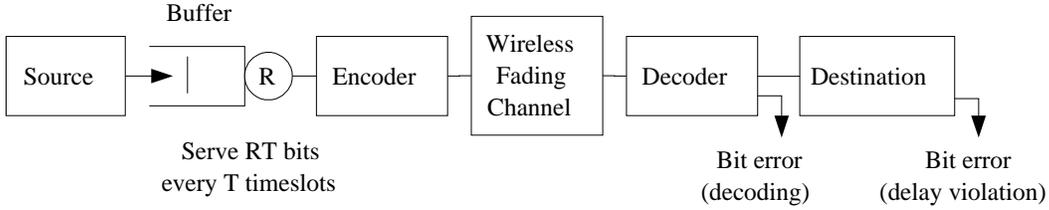}
\caption{System model} \label{fig:ModelDiversity}
\end{center}
\vspace{-0.1in}
\end{figure*}

This work analyzes the high signal-to-noise-ratio (SNR) performance of outage-limited communications where the information to be communicated is delay-limited and where the
information arrives at the transmitter in a stochastic manner. We consider the following setting (Figure~\ref{fig:ModelDiversity}) in our study:

\bit
\item A random number of bits arrive at the transmitter during any given timeslot. Bits accumulate in an infinite buffer while waiting for their turn to be bunched into codewords and transmitted under a first-come, first-transmit policy.

\item There is no feedback to the transmitter; retransmission of the bits in error is not considered.

\item Communication over the fading channel is outage-limited
(\!\cite{ZheTse,OzarowShamaiWyner1994}), 
where the transmitter is unaware of the instantaneous channel state and, as a consequence, operates at a fixed transmission rate, $R$. During a deep fade (also known as an outage), the channel seen by the decoder is too weak to allow recovery of the data content from the transmitted signal.  Characteristic settings are those of MIMO and cooperative outage-limited communications.

\item Coding takes place in blocks where each codeword spans over a fixed and finite integral number, $T$, of timeslots. Each codeword has an information content of $RT$ bits. In addition, coding is ``fully-diverse,'' i.e., the decoding at the receiver takes place only at the end of the coding block.

\item The delay bound, $D$, is a maximum allowable time duration from the moment a bit arrives at the transmitter until the moment it is decoded at the receiver. The delay experienced by a bit is the sum of the time spent waiting in the buffer and the time spent in the block decoding process. Note that the waiting time in the buffer is random due to the stochastic arrival process.

\item A bit is declared in error either when it is decoded incorrectly at the decoder, or when it violates the delay bound.
\eit

For the above setting, we are interested in the high-SNR asymptotic total probability of bit error. 
Note that for a given transmission rate, $R$, and a coding block duration, $T$, there exists a tradeoff between the probabilities of decoding error versus the delay violation.
We expect that longer coding blocks allow the encoded bits to be transmitted over more fading realizations and hence, achieve higher diversity and fewer decoding errors. However, longer coding blocks cause more bits to violate the delay requirement. 
In other words, one intuitively expects that there is an optimal choice of the fixed transmission rate, $R$, and the fixed coding block duration, $T$, for which the total probability of bit error is minimized. The goal of this paper is to analytically identify these optimal quantities.

\subsection{Prior Work and Our Contribution} 
High demands on the quality of service (QoS), in terms of both
packet losses and packet delays, have fueled substantial research
interest in jointly considering channels and queues. Communication
of delay-sensitive bits over wireless channels has been addressed under various assumptions and settings
in works such as
\cite{BerryLargeDelay,BerrySmallDelayITA06,RajanSabharwalAazhang04,Negi,BetteshShamai,LiuParag}. Often,  asymptotic approximations are employed to enable tractable analysis of the problem. Below we detail the existing work with their corresponding settings and the relationships to this paper.

The first group we discuss, \cite{BerryLargeDelay,BerrySmallDelayITA06,RajanSabharwalAazhang04,Negi}, consists of scenarios where the current channel state information (CSI) is assumed to be known at both the transmitter and receiver. For example, in \cite{BerryLargeDelay} and \cite{BerrySmallDelayITA06}, Berry and Gallager address the tradeoff between the minimum average power consumption and the average delay (the power-delay tradeoff) over a Markovian fading channel with CSI both at the transmitter and the receiver. In such a setting, the transmitter dynamically varies power (i.e., the rate) in response to the current queue length and channel state. In \cite{RajanSabharwalAazhang04}, Rajan et al. derive optimal delay-bounded schedulers for transmission of constant-rate traffic over finite-state fading channels.
In \cite{Negi}, Negi and Goel apply the effective capacity \cite{EffectiveCapac} and error exponent
techniques to find the code-rate allocation that maximizes the decay rate of the asymptotic probability
of error for a given asymptotically-large delay requirement.
Similar to \cite{BerryLargeDelay} and \cite{BerrySmallDelayITA06}, the proposed dynamic code-rate allocation in \cite{Negi} is in response to the current channel fading and is possible by assuming CSI knowledge at the transmitter.

A second group of work (e.g., \cite{BetteshShamai,LiuParag}) focuses on scenarios where CSI is unknown to the transmitter but there is a mechanism for retransmission of codewords when the channel is in outage.
As a tradeoff to protection against channel outage,
this retransmission incurs extra delays to the bits in the buffer. In \cite{BetteshShamai}, for example, Bettesh and Shamai (Shitz) address the problem of minimizing the average delay, under average power constraints and fixed transmission rate. They provide asymptotic analysis, under heavy load condition and asymptotically large queue length, for the optimal adaptive policies that adjust the transmission rate and/or transmission power in response to the current queue length at the transmitter.
In another example, Liu et al. in \cite{LiuParag} study the problem of optimal (fixed) transmission rate to maximize the decay rate of the probability of buffer overflow for on-off channels and Markov-modulated arrivals. The channel is considered ``off'' when outage occurs.

Although our work uses a similar performance measure to \cite{Negi}, namely the decay rate of the asymptotic probability of error, it covers the scenarios in which CSI is not available to the transmitter (no CSIT) and there is no retransmission. In such a setting, the variation of the fading channel is combatted via a coding over multiple independent fading realizations.\footnote{For example, the multiple independent fading realizations can be a result of fading in multiple channel coherence time intervals (known as time diversity), or fading in multiple independent spatial channels, as in MIMO channel (spatial diversity), or cooperative relay channel (cooperative diversity).} While this approach improves the transmission reliability, its longer coding duration increases the end-to-end delay any bit faces,  and can potentially increase the probability of delay violation. In other words, in the absence of CSIT and retransmission, the transmission reliability, as well as the delay violation probability, are functions of the coding rate and duration. Consequently, our work compliments this previous research as it considers the effect of a delay violation requirement, 
in the absence of CSI at the transmitter and retransmission, on the operation of the physical layer. We consider a fixed transmission rate and code duration, as opposed to dynamic policies.

Since it is difficult to derive the exact relationship between the system parameters and the probabilities of channel decoding error and the delay violation, we choose to study an asymptotic approximation when the signal-to-noise ratio (SNR) is asymptotically high. The first advantage of this choice is the availability of an asymptotic high-SNR analysis for the channel decoding error probability. This high-SNR analysis is known as the \emph{diversity-multiplexing-tradeoff} (DMT) analysis~\cite{ZheTse} and has received a great deal of attention during the past few years. Another advantage of the high-SNR analysis is that, for the class of arrival processes we consider in this paper, we can derive an asymptotic approximation of the delay violation probability that is valid even when the delay requirement $D$ is finite and small. This derivation (Lemma~\ref{lem:Pdv}) is based on a large-deviations result known as the G\"artner-Ellis theorem (see e.g., \cite{Dembo}) and extends the large deviations exponent for a queue with asymptotic number of flows (as provided in \cite{Weiss86,BotDuf95,CouWeb96,BigQueues}) to a queue with batch service discipline.
Given that the asymptotic expression of the total probability of bit error is valid without requiring asymptotically large $D$, it is then meaningful to ask about the optimal coding block duration, a question which is not answered in studies with asymptotic $D$ (e.g., \cite{BerryLargeDelay,Negi,MIMOMac,Somsak,EliKitJav_WiOpt_2007,BerrySmallDelayITA06}).



We also would like to point out that our work was motivated by the work of Holliday and Goldsmith \cite{Goldsmith} where, under a high-SNR asymptotic approximation, the optimal operating channel transmission rate for a concatenated source/channel system is studied. Following the approach in \cite{Goldsmith}, we study a concatenated queue/channel system under a high-SNR approximation.

\subsection{Overview of the Results}
This work 
focuses on the notion of SNR error exponent as a measure of performance. 
Specifically, we are interested in finding how the asymptotic total probability of error decays with SNR. To keep the problem meaningful, we consider a scenario under which the overall traffic loading of the system (the ratio between the mean arrival rate and the ergodic capacity of the channel) is kept independent of SNR. That is, we consider a case where the arrival rate scales with $\log \SNR$.  Note that this scaling of arrival process is necessary to ensure a fixed loading and hence a comparable cross-layer interaction as SNR scales.

From the DMT result, we already know that, if the channel operates below the channel ergodic capacity, the asymptotic probability of channel decoding error decays with SNR. The best one can hope for is that the asymptotic total probability of error decays exponentially with SNR. For that, the asymptotic probability of delay violation needs to decay with SNR. 
Specifically, we consider a class of i.i.d.\footnote{\ska Note that, 
since the adopted channel model is not assumed to be i.i.d., assuming an i.i.d. arrival process, intuitively, is not consequential: think of our chosen time slot as an upperbound for the ``coherence time'' of the arrival process. The i.i.d. source assumption mostly serves to simplify the exposition and presentation of results, and does not fundamentally limit the setting. \skb} arrival processes with light tail (i.e., the processes have all moments finite) whose burstiness (defined as the ratio of the standard deviation over the mean of the number of bits arrived at a timeslot) monotonically goes to zero as SNR goes to infinity. We show that for all such processes (called smoothly-scaling processes), the total probability of error decays.

The main result of the paper shows that the optimal decay behavior of the asymptotic total probability of bit error depends on how fast the burstiness of the source scales down with SNR. If the source burstiness scales down too slowly (too quickly), the majority of the errors are due to delay violation (channel error), i.e., the total probability of error is asymptotically dominated by delay-violation (channel-error) events.
However, at the proper scaling where the burstiness scales linearly with $\frac{1}{\sqrt{\log\SNR}}$ and with the optimal coding duration and transmission rate, the occurrences of channel errors and delay-violation errors are asymptotically balanced. Equivalently, one can interpret our result,
the optimal choice of block coding duration and transmission rate, as that which balances the channel atypicality (deep fading or outage events) and the arrival atypicality (large bursts of arrivals).

We apply this result to several examples of outage-limited communication systems to find the optimal setting of the operating parameters.

\subsection{Outline of the Paper}
The precise models for the coding/channel process and the bit-arrival/queue process are described in Section \ref{sec:system Model}. We precisely define the scaling of the source process with SNR and give a simple example of such source processes. Section~\ref{sec:Pdv} provides the asymptotic probability of delay violation. The main result of the paper is found in Theorem~\ref{thm:1} of Section~\ref{sec:RDMT_GeneralSetting}. This theorem provides the optimal asymptotic decay rate of the total error probability as well as the optimal coding duration and transmission rate. 
Section~\ref{sec:discussionOfResults} gives some examples to illustrate the utility of Theorem~\ref{thm:1}.
These examples consider the question of optimally communicating delay sensitive packet stream with a compound Poisson traffic profile over the following outage-limited channels: SISO Rayleigh fast-fading channel, quasi-static cooperative relay channel, and quasi-static MIMO channel.
Section~\ref{sec:conclussionWiopt} concludes the paper. Appendices include the proofs of various lemmas and theorems.

\subsection{Notations}
We use the following symbols and notations. We use $\rho$ to denote SNR.
The notation $\eqg$ for a strictly increasing and positive-valued function $g$ represents the equivalence between $y(\rho) \eqg z(\rho)$ and $\lim\limits_{\rho\rightarrow \infty} \frac{\log y(\rho)}{g(\log \rho)} = \lim\limits_{\rho\rightarrow \infty} \frac{\log z(\rho)}{g(\log \rho)}$.
We define $\geg$ and $\leg$ in a similar manner. Note that when $g$ is an identity function, then $\eqg$ is equivalent to the familiar $\doteq$ notation in the DMT analysis \cite{ZheTse}.

We denote the high-SNR approximation of the ergodic capacity of AWGN channel by $N:=\log \rho$ and use $N$ and $\log \rho$ interchangeably.  The sets $\Z$, $\N$, and $\No$ represent the set of all, positive, and non-negative integers, respectively. In addition, the set $\T$ represents the set $\{1,2,\ldots,\left\lfloor \frac{D}{2}\right\rfloor \}$. 
Flooring and ceiling functions are denoted by
$\lfloor \cdot \rfloor$ and $\lceil \cdot \rceil$, respectively. For all $a\le b$, $[x]_a^b = \max(a,\min(b,x))$ and $[x]^+ = \max(x,0)$. 
We write $g(x) = \Theta(h(x))$ to denote that the function $g$ scales linearly with the function $h$, i.e.,  $\lim\limits_{x\rightarrow \infty} \frac{g(x)}{h(x)} < \infty$ and $\lim\limits_{x\rightarrow \infty} \frac{h(x)}{g(x)} < \infty$. Finally, for any function $f$, we denote its convex conjugate, $f^*$, by \beq f^*(x) = \sup_{\theta\in \R} \ \theta x - f(\theta).\eeq

\section{System Model\label{sec:system Model}}
As discussed in the introduction, we consider a system composed of a bursty and delay-limited information source, concatenated with an infinite buffer and a fading channel, as shown in Figure~\ref{fig:ModelDiversity}. We assume the queue follows a first-come-first-serve (FCFS) discipline. The departures out of the queue occur according to a block channel coding scheme, while the arrivals to the queue follow a stochastic model. If the transmission rate is above the instantaneous capacity of the channel, an outage event is said to occur where the received signal is erroneously decoded.
The delay requirement asks that each bit of information be decoded at the destination within a \emph{maximum allowable delay} of $D$ time-slots from the time it arrives at the buffer. Otherwise, the bit will be obsolete, discarded, and counted as erroneous.We assume no retransmission of unsuccessful transmissions or those bits which violate the delay bound.\footnote{Note that due to the constant service rate of the queue and the FCFS service discipline, any bits arriving at the queue know immediately whether they will exceed their delay constraints, using the knowledge of the current queue length. It seems wise to drop these bits immediately after their arrivals to improve the system performance. However, we do not need to consider such method because it has been established (see \cite[Theorem~7.10]{BigQueues}) that, in the asymptotic regime of interest, such method does not improve the exponent of the delay violation probability.} In the next three subsections, we describe in detail the models for the channel, the arrival process, and the system performance measure.

\subsection{Channel and Coding Model}
We consider a general fading-channel model,
$$ \underline{y} = H \underline{x} + \underline{w},$$
where $\underline{x}$ is the transmitted vector, $H$ is the channel matrix, $\underline{y}$ is the received signal, and $\underline{w}$ is the noise vector. 
The average SNR is defined as \cite{ZheTse}
\[ \rho:=\frac{\mathbb{E}[\| H \underline{x} \|^2
]}{\mathbb{E}[\|\underline{w}\|^2]}, \]
and in the asymptotic scale of interest, 
it is equivalent to \[\rho \doteq
\mathbb{E}[\|\underline{x}\|^2].\]
Coding takes place over $T$ timeslots, using rate-$R$,
length-$T$ codes that meet the DMT
tradeoff $\dch(r,T)$ \cite{ZheTse}, defined as \beq \label{eq:DMT} \dch(r,T) := -\limrho
\frac{\log \Pch(r,T,\rho)}{\log\rho}, \eeq where $P_{\text{ch}}(r,T,\rho)$ is the codeword error probability induced by the channel, given an
optimal code of \emph{multiplexing gain} $r$, coding block size $T$ timeslots\footnote{For most settings, there exist codes
that meet the entire DMT tradeoff in minimum delay, independent of
channel dimensionality and fading statistics \cite{EliKumPawKumLu,Belfiore_Dynamic_AAF_journal,TavVisUniversal_2005}.}, and average SNR $\rho$. The channel multiplexing gain $r$ is related to the transmission rate $R$ as (refer to \cite{ZheTse}) \beq r := \limrho \frac{R}{\log \rho}.\eeq That is, the transmission rate $R$ is assumed to scale linearly as $r \log \rho$. We denote by $r_\text{max}$ the maximum value of $r$, i.e., $0\le r \le \rmax$. This $r_{\max}$ relates to the ergodic capacity as \[  r_{\max} =
\mathbb{E}_{H} \frac{ \max_{p_x} I(\underline{x};\underline{y})  }{
\log\rho  }  \]
and is the smallest $r$ such that $d_{\text{ch}}(r,T)=0$.

The DMT tradeoffs have been extensively studied for various finite-duration communication schemes (for example, see  \cite{GamCaiDam,TavVisUniversal_2005,EliKumPawKumLu,CaiEliRaj,EliSetKum}
for MIMO point-to-point communications, \cite{TseVisZhe_MultAccess_2004} for multiple access
communications, 
\cite{LanWorTSEIEEE,AzaGamSch} for cooperative
communications, and \cite{Petros_Thesis,Belfiore_Dynamic_AAF_journal}
for cooperative communications with small delay).

\begin{note} The condition that each bit be transmitted over all
timeslots in the coding block\footnote{Currently, all known minimum-delay DMT optimal
codes over any fading channel with non-zero coefficients ask that
each bit be transmitted over each timeslot.}, together with the
first-come first-transmit service discipline, makes it so that every $T$
timeslots, the $RT$ oldest bits are instantaneously removed\footnote{If an insufficient number of bits exists in the buffer, null bits are used and the rate is maintained. It is easy to show that,
in the asymptotic scale of interest, the use of null-bits does not
incur any change in the SNR exponent of the probability of error. } from
the queue and are transmitted over the next $T$ timeslots. We assume that it is only at the end of the $T$ timeslots that all the $RT$ bits are decoded by the decoder.
\end{note}

\begin{ex}[Rayleigh Fast-Fading SISO Channel]
Consider the single-input single-output (SISO) time-selective channel with Rayleigh
fading coefficients (correlated or uncorrelated) and with additive white
Gaussian noise at the receiver.
The corresponding channel model over $T$ timeslots is given by
\[\underline{y} = \text{diag}(\underline{h}) \ \underline{x}+\underline{w},\] where  $\underline{y},\underline{h},\underline{x},$ and $\underline{w}$ are $T\times 1$ vectors and  $H = \text{diag}(\underline{h})$ is a $T \times T$ diagonal fading matrix with the fading at the $t$th timeslot, $h_t$, as its $(t,t)$ element.  The optimal DMT, given optimal signaling, takes the form
\bean
d_{\text{ch}}(r, T ) &:= & -\limrho \frac{\log \text{Pr}\bigl(
I(\underline{x};\underline{y} | \underline{h}  )  <2^{RT}\bigr)
}{\log\rho} \no \\
&= & -\limrho \frac{\log \text{Pr}\bigl(\prod_{t=1}^T
(1+\rho|h_t|^2)<\rho^{rT}\bigr)}{\log\rho}.  \no\\
\eean
For the fast-fading case where the coherence time is equal to one
timeslot and the elements of $\underline{h}$ are Rayleigh i.i.d. random
variables, the tradeoff takes the form
\beq d_{\text{ch}}(r, T ) = T(1-r), \label{eq:DMTfastFading} \eeq
and it can be met entirely in T timeslots (see [1]). This SISO channel
allows for $r_{\max} = 1.$

\end{ex}

Other examples which will be discussed later in Section~\ref{sec:discussionOfResults} are quasi-static MIMO and cooperative-relay channels. 
In this paper, for simplicity we assume that $\dch(r,T)$ is continuous on $r$, decreasing on $r$, and increasing on $T$.
\subsection{Smoothly-Scaling Bit-Arrival Process} 
In this subsection, we describe the SNR-scaling of a family of arrival processes of interest. The specific choice of SNR-scaling for the statistics of the bit-arrival process is such that the average traffic load of the system (defined as the ratio of the average arrival rate over the ergodic capacity) is kept constant, independent of SNR.\footnote{It can be seen that unless the traffic load (average bit arrival rate over the channel rate) scales as $\log(\SNR)$, i.e., $\limrho \frac{\mathbb{E}[A_t]}{\log\rho} = \ell$ for some fixed $0<\ell<\infty$, the problem is void of cross-layer interactions. Otherwise if $\ell=0$, corresponds to the case where too few bits arrive and effectively there is no queuing delay.
On the other hand, when the traffic load scales much faster than $\log(\SNR)$, i.e., $\ell=\infty$, the overall performance is dominated by queueing delay, independently of the channel characteristics.}
This means that scaling in the ergodic capacity $\rmax \log\rho$ ($=\!\rmax N$) is matched by scaling the average bit-arrival rate as  $\lambda\log \rho$ ($=\!\lambda N$) as well, for some $\lambda >0$. Now we are ready to introduce the arrival process of interest: The sequence of asymptotically \emph{smoothly-scaling}
bit-arrival processes, in which the process becomes ``smoother'' for increasing $N$.

\begin{defn} Let $\G$ denote a class of functions which contains any function
$g:\R^+ \mapsto \R^+$ (called \emph{scaling function}) which is continuous and strictly increasing and whose tail behavior is such that 
\beq \lim\limits_{x\rightarrow \infty} \frac{g(x)}{\log x} = \infty.\label{gscaling}\eeq
\end{defn}

\begin{defn} \label{defn1} (\emph{$g$-smoothly-scaling source})
Consider a scaling function $g\in\G$ and a family of bit-arrival processes $(A^\Ng,N \in \N)$, where  
$A^\Ng = (A^\Ng_t, t\in\Z)$ denotes an i.i.d. sequence of the random numbers $A^\Ng_t$ of bits that arrive at time $t$ with
$E[A^\Ng_t] = \lambda N$, for all $t$. The family of bit-arrival processes is said to be
\emph{$g$-smoothly-scaling} if the \emph{limiting $g$-scaled logarithmic moment generating function}, defined for each $\theta \in \R$ as
\beq \Lambda(\theta) = \lim_{N\rightarrow \infty} \frac{\log E[\exp(\frac{\theta g(N)}{N}A^\Ng_1)]}{g(N)}, \label{Gen:condA} \eeq
exists as an extended real number in $\R^*:=\R\cup\{\infty\}$ and is finite in a neighborhood of the origin, \emph{essentially smooth}, and \emph{lower-semicontinuous}.\footnote{\cite{BigQueues} A function $f:\R \mapsto \R^*$ is \emph{essentially smooth} if the interior of its effective domain $\mathcal{D} = \left\{x:f(x) < \infty\right\}$ is non-empty, if it is differentiable in the interior of $\mathcal D$ and if $f$ is steep, which means that for any sequence $\theta_n$ which converges to a boundary point of $\mathcal D$, then $\lim_{n\rightarrow \infty} |f'(\theta_n)| = +\infty$. $f$ is \emph{lower-semicontinuous} if its level sets $\left\{x: f(x) \le \alpha\right\}$ are closed for $\alpha \in \R$.}
\end{defn}

\begin{note} \label{rem:Lambda}
It is straight forward to show that $\Lambda$ is convex and $\Lambda'(0) = \lambda$ (see \cite[Lemma~1.11]{BigQueues}).
\end{note}

Note that $\lambda$ describes how close the average bit-arrival rate is to the asymptotic approximation of the ergodic capacity of the channel.  For stability purpose and to ensure the existence of a stationary distribution, we require that $\lambda < \rmax$.
Also, note that we abuse the notation and denote the arrival process by $A^\Ng_t$, despite its possible dependency on the scaling function $g$.

\subsubsection{Motivation for Smoothly-Scaling Assumption}
The assumption of $g$-smoothly-scaling arrival processes allows us to find the decay rate of the tail probability of the sequence of process $(S^\Ng_t,N\in\N)$, which is a sum process
defined as $$S^\Ng_t = \sum_{j=1}^t A^\Ng_j, \ \ t\in \N;$$
since $(A^\Ng_j,j\in\Z)$ are i.i.d., $S^\Ng_t$
is also a $g$-smoothly scaling process with the limiting $g$-scaled log moment generating function $\Lambda_{S_t}$ given as 
\beq \Lambda_{S_t}(\theta) := \lim_{N\rightarrow \infty} \frac{\log E[\exp(\frac{\theta g(N)}{N} S^\Ng_t)]}{g(N)}  = t\Lambda(\theta)
\label{Gen:condS}. \eeq 
Now, given that the sequence $(S^\Ng_t,N\in\N)$ is $g$-smoothly-scaling, we can use the G\"artner-Ellis theorem (see e.g., \cite{Dembo} and \cite{BigQueues}) 
to give the following result on the decay rate of the tail probability of the sequence. The following proposition provides an important basis for the analysis of the asymptotic probability of delay violation in Section~\ref{sec:Pdv}.

\begin{prop}\emph{(G\"artner-Ellis theorem for $g$-smoothly-scaling process)}  \label{GartnerEllisSmoothlyScaling}
Consider $g\in \G$ and a family of $g$-smoothly-scaling processes $(A^\Ng, N\in \N)$
with the limiting $g$-scaled log moment generation function $\Lambda$. Let $S^\Ng_t= \sum_{i=1}^t A^\Ng_i$, for $t\in \N$.
Then, for $a>\lambda t$, we have \beq
\lim_{N\rightarrow \infty} \frac{1}{g(N)} \log P\left(\frac{S^\Ng_t}{N} > a\right) = -t\Lambda^*(a/t),\label{eq:Prop1} \eeq where $\Lambda^*$ is the convex conjugate of $\Lambda$.
\end{prop}
\begin{proof} See Appendix~\ref{append:pffact3}.
\end{proof}

\subsubsection{Asymptotic Characteristic of Smoothly-Scaling Processes} \label{subsect:AsympChar}
 Intuitively, the $g$-smoothly-scaling arrival processes become smoother as SNR increases.
This intuition follows from \reqn{Gen:condA}, which implies that for $\theta\in \R$ such that $\Lambda(\theta) < \infty$ and $\epsilon > 0$, there exists $N_0$ such that for $N>N_0$,
\begin{align*} \exp\left(g(N) \Lambda(\theta) - g(N)\epsilon\right) &< E\left[\exp\left(\frac{\theta g(N)}{N}A^\Ng_1\right)\right] \\ &< \exp\left(g(N) \Lambda(\theta) + g(N)\epsilon\right). \end{align*}
 Then, if we let $Y_{g(N)}$ be a sum of $g(N)$ i.i.d. random variables (i.e., $Y_{g(N)} := X_1+\cdots+X_{g(N)}$ with $E[e^{\theta X_1}] = e^{\Lambda(\theta)}$), we have  $E[e^{\theta(Y_{g(N)})}] = e^{\Lambda(\theta) g(N)}$. Therefore, at sufficiently large $N$, $\frac{g(N)}{N} A^\Ng_t$ and $ Y_{g(N)}$ have the same moment generating function and hence the same distribution.
If we define the burstiness of the random variable $A^\Ng_1$ as the (dimensionless) ratio of its standard deviation over its mean,\footnote{Note that the burstiness definition here is basically the normalized variation of the random variable around its typical value (its mean).
A more familiar definition of traffic burstiness would involve how the traffic are correlated with time, i.e., a bursty source tends to have large bursts of arrivals in a short period of time. However, since we only consider the source which is i.i.d. over time, we use this definition of burstiness.
} then, using the above intuition, the burstiness $\frac{\text{std}(A^\Ng_1)}{E[A^\Ng_1]} $ for large $N$ is approximately equal to $\frac{\text{std}\left(\frac{N}{g(N)} \sum_{i=1}^{g(N)} X_i\right)}{\lambda N}$, which is reduced to $\frac{\text{std}(X_1)}{\lambda \sqrt{g(N)}}$.
Hence, the burstiness of $A^\Ng_1$
decays to zero approximately as $\frac{1}{\sqrt{g(\log \rho)}}$.
In other words, the $g$-smoothly-scaling arrival processes become smoother as SNR increases.

\subsubsection{Examples of Smoothly-Scaling Processes}
One of the common arrival processes used for traffic modeling is
a \underline{c}ompound \underline{P}oisson process with \underline{e}xponential packet size, denoted as CPE. For this source, the random number of bits, $A^\Ng_t$, arrived at timeslot $t$, is i.i.d. across time $t$ and is in the form of \beq A^\Ng_t =
\sum_{i=1}^{M_t^\Ng} Y_{i,t}^\Ng, \eeq where $M_t^\Ng$ is the random
variable corresponding to the number of packets that have arrived
at the $t^{th}$ timeslot, and where $Y_{i,t}^\Ng$ corresponds to
the random number of bits in the $i^{th}$ packet. $M_t^\Ng$ are
independently drawn from a Poisson distribution with mean
$\nu(N)$; and $Y_{i,t}^\Ng$, $i=1,\ldots, M_t^\Ng$, are independently drawn from an exponential distribution with mean $\frac{1}{\mu(N)}$ (nats per packet). Note that the assumption that $E[A^\Ng_t] = \lambda N$ forces that $\frac{\nu(N)}{\mu(N)} = \lambda N$. In addition, a larger average packet size $\frac{1}{\mu(N)}$ implies a more bursty arrival process.\footnote{It can be easily shown that the burstiness of this CPE process, as defined in Section~\ref{subsect:AsympChar}, is $\sqrt{\frac{2}{\lambda N \mu(N)}}$.}
It is known (see \cite{Somsak}) that the log moment generating function of this CPE random variable $A^\Ng_t$ is \beq \log E[e^{\theta A^\Ng_t}] = \begin{cases} \frac{\theta \nu(N)}{\mu(N)-\theta}, &\theta < \mu(N),\\ \infty,  &\text{ otherwise.}\end{cases}\label{MGFCPE} \eeq

The following examples illustrate that, depending on the scaling of the average packet arrival rate and the average packet size, some CPE processes may or may not be $g$-smoothly-scaling.

\begin{ex}[$g$-smoothly-scaling CPE process] \label{example2} For $g\in\G$ and $\mu>0$, consider a CPE process $A^\Ng_t$ with packet arrival rate $\mu \lambda g(N)$ and average packet size $\frac{N}{\mu g(N)}$. This family of processes is $g$-smoothly-scaling because, using (\ref{MGFCPE}), we have \beq \Lambda(\theta) := \limN \frac{\log E[e^{\frac{\theta g(N)}{N} A^\Ng_1}]}{g(N)} = \begin{cases} \frac{\mu\lambda\theta}{\mu-\theta}, &\theta < \mu,\\ \infty, &\text{ otherwise},\end{cases} \label{defn:LambdaCPE} \eeq which satisfies the conditions in the definition of $g$-smoothly-scaling.
Since we will use this particular $g$-smoothly-scaling CPE process for examples in the paper, we denote it as CPE$(\lambda,\mu,g,N)$.
 It is useful to note a particular case when $g(N)$ grows linearly with $N$. Using a property of the Poisson process \cite{Viniotis98}, this particular scaling case can be viewed as aggregating an increasing number of 
Poisson traffic streams (this number grows linearly with $N$), with each stream having the same packet length distribution. 
\end{ex}

To complete our discussion on smoothly-scaling processes, we give an example below of a family of CPE arrival processes which is not $g$-smoothly-scaling.

\begin{ex}  A family of CPE processes where $A^\Ng_t$ has packet arrival rate $\mu\lambda$ and average packet size $N/\mu$ (note the dependence on $N$ only in the average packet size) is not $g$-smoothly-scaling for any $g\in\G$. This is because, using (\ref{MGFCPE}), we have $$\limN \frac{\log E[e^{\frac{\theta g(N)}{N} A^\Ng_t}]}{g(N)} = \begin{cases} 0, &\theta \le 0,\\\infty  &\text{ otherwise,}\end{cases}$$ which is not finite in the (open) neighborhood of $\theta=0$. Hence, this family of processes is not $g$-smoothly-scaling. 
\end{ex}

\begin{note} \ska The scaling function, $g$, describes the way the source statistics scale with SNR.  Example~\ref{example2} describes the case of the compound Poisson process, where $g$ can be identified as the function that specifies how the average packet arrival rate ($\mu \lambda g(\log \SNR)$) and the average packet size ($\frac{\log \SNR}{\mu g(\log \SNR)}$) scale with SNR. \skb
\end{note}



\subsection{Performance Measure and System Objective}
The overall performance measure is the total probability of bit loss, $\Pt(r,T)$, where loss can occur due to channel decoding error or the end-to-end delay violation. Specifically,
\beq \Pt(r,T) := \Pch(r,T) + \left(1-\Pch(r,T)\right)\Pdv(r,T),\label{defn:Ptot} \eeq
where $\Pch(r,T)$ denotes the probability of decoding error due to channel outage and $\Pdv(r,T)$ denotes the probability of delay violation. We are interested in finding the high-SNR asymptotic approximation of $\Pt(r,T)$ as a function of $r$, $T$, SNR, $D$, as well as the source and channel statistics (including $\lambda$ and the source scaling function $g$). In the interest of brevity, we denote $\Pt$ as a function of only $r$ and $T$, the two parameters over which the performance will later be optimized.


Since the high-SNR asymptotic expression of $\Pch(r,T)$ is already given by the DMT in (\ref{eq:DMT}), what remains is to find the asymptotic expression for $\Pdv(r,T)$, which is shown in the next section.

\section{Asymptotic Analysis of Probability of Delay Violation} \label{sec:Pdv}
In this section, we derive the asymptotic probability of delay violation $\Pdv(r,T)$ for the channel multiplexing rate $r$ and coding block size $T$. 
 We observe that the adopted block coding forces the queue to have a \emph{batch service} that occurs every $T$ timeslots with the instantaneous removal of the oldest $rNT$ bits. The decay rate of the asymptotic tail probability of the sum arrival process, given in  Proposition~\ref{GartnerEllisSmoothlyScaling}, in conjunction with an asymptotic analysis of a queue with deterministic batch service, gives the following result:

\begin{lem} \label{lem:Pdv}
Given $g\in\G$, $T \in \T$, $r>\lambda$, a batch service of $rNT$ every $T$ timeslots, and a $g$-smoothly-scaling bit-arrival process characterized by the limiting $g$-scaled log moment generation function $\Lambda$, the decay rate of $P_\text{delay}(r,T)$ is given by the function $I$, i.e., \beq
\limN \frac{1}{g(N)} \log P_\text{delay}(r,T) = -I(r,T),\eeq
where \begin{align} &I(r,T) \no\\&= 
\min_{\substack{t\in\No:\\tT+T-1-k>0}} \ (tT+T-1-k) \Lambda^*\left(r+\frac{(D+1-2T)r}{tT\!+\!T\!-\!1\!-\!k}\right),\label{defn:I}\end{align}
for $k = D(\!\!\!\mod T)$. In addition, 
 $I(r,T)$ is lower-semicontinuous and increasing on $r$. 
\end{lem}
\begin{proof} See Appendix~\ref{appen:PfPdv}. \end{proof}


\begin{approximation} 
\label{ApproxI}
Relaxing the integer constraint in (\ref{defn:I}) gives the lower bound of $I$ as \beq I(r,T) \ge \delta_r r (D+1-2T) =: I_{ir}(r,T),\eeq where \beq \delta_r = \sup\{\theta >0: \Lambda(\theta) < \theta r\}.\label{defn:deltar} \eeq We use this lower bound as an approximation to $I$ as well, i.e., \beq I(r,T) \approx I_{ir}(r,T) = \delta_r r (D+1-2T).\label{I:approx} \eeq
\end{approximation}
\begin{proof} See Appendix~\ref{appen:PfPdv}. \end{proof}

\begin{ex} For a $g$-smoothly-scaling CPE($\lambda,\mu,g,N$) bit-arrival process, the function $I$ in (\ref{defn:I}) can be calculated exactly with the following $\Lambda^*$: 
\beq \Lambda^*(x) = \mu\left(\sqrt{x}-\sqrt{\lambda}\right)^2, \ \ x\in\R.\eeq
However, an approximation of $I$ in (\ref{I:approx}) is simpler to work with and given as \beq \Icpe(r,T) \approx I_{ir}(r,T) = \mu(r-\lambda) (D+1-2T),\label{approxPdvCPE} \eeq where, using (\ref{defn:deltar}) and (\ref{defn:LambdaCPE}), $\delta_r$ is given as \beq \delta_r = \mu \left(1-\frac{\lambda}{r}\right).\label{eq:delta11}\eeq
We will see via numerical examples in Section~\ref{subsec:NumComparison} that the approximation in (\ref{approxPdvCPE}) is sufficient for our purpose.
\end{ex}

\section{Main Result: Optimal Asymptotic Total Probability of Error}
\label{sec:RDMT_GeneralSetting}
In this section, we present the main result of the paper which states the optimal decay rate of the high-SNR asymptotic total probability of bit error. Recall the definition of $\Pt$ from (\ref{defn:Ptot}): $$\Pt(r,T) := \Pch(r,T)+(1-\Pch(r,T)) \Pdv(r,T),$$
where we now know that $$\Pch(r,T) \doteq \rho^{-\dch(r,T)}$$ and $$\Pdv(r,T) \eqg e^{-I(r,T) g(\log\rho)}.$$
Hence, the asymptotic optimal decay behavior of $\Pt$ depends on the function $g$.
The following theorem gives the main result of the paper.

\begin{thm} \label{thm:1}
Consider $g\in\G$ and a $g$-smoothly-scaling bit-arrival process. The optimal rate of decay of the asymptotic probability of total bit error, maximized over all $r \in (\lambda,\rmax)$ and $T\in \T$, and the optimizing $r^*$ and $T^*$ are given, depending on the tail behavior of the function $g$, as follows:

\underline{Case 1}: If $\lim\limits_{N\rightarrow \infty} \frac{g(N)}{N} = \gamma \in (0,\infty)$, then  \bea  d^* &:=& \sup_{\substack{r\in (\lambda,r_\text{max})\\T \in \T}} \ \limrho \frac{-\log \Pt(r,T)}{\log \rho} \no\\ &=& \dch(r^*,T^*) = \gamma I(r^*,T^*),\label{eq:FirstPart1}\eea where \bea
r^*(T) &:=& \inf\{r\in (\lambda,\rmax): \gamma I(r,T) = \dch(r,T)\}\\ 
T^* &=& 
\arg\max\limits_{T\in \T} \  I(r^*(T),T)  \label{main:T} \\ r^* &= & r^*(T^*). \label{eq:FirstPart4}\eea

\underline{Case 2}: If $\lim\limits_{N\rightarrow \infty} \frac{g(N)}{N} = 0$ and $\lim\limits_{N\rightarrow \infty} \frac{g(N)}{\log N} = \infty$, then
\beq \sup_{\substack{r\in (\lambda,\rmax),\\T \in \T}} \ \limrho \frac{-\log \Pt(r,T)}{g(\log \rho)} \le  \max_{T \in \T} \ I(\rmax,T).\eeq

\underline{Case 3}: If $\lim\limits_{N\rightarrow \infty} \frac{g(N)}{N} = \infty$, then
\beq \sup_{\substack{r\in (\lambda,\rmax),\\T \in \T}} \ \limrho \frac{-\log \Pt(r,T)}{\log \rho} \le \dch\left(\lambda,\left\lfloor \frac{D}{2}\right\rfloor\right). \hspace{.2in} \eeq  
\end{thm}
\begin{proof} See Appendix \ref{sect:ProofGammaConst}.
\end{proof}

\

Theorem~\ref{thm:1} shows that the optimal decay behavior of the asymptotic total probability of error depends on the tail behavior of the function $g$. As discussed earlier, the burstiness of the $g$-smoothly-scaling arrival process scales down as $\Theta(\frac{1}{\sqrt{g(\log\rho)}})$. Below, we discuss each case of Theorem~\ref{thm:1}, with respect to the scaling of the source burstiness:

In Case 1, where the source burstiness scales down with $\Theta(\frac{1}{\sqrt{\log\rho}})$, 
both components of the probability of error decay exponentially with $\log \rho$. In this setting, one can optimize the choices of $r$ and $T$ to arrive at a non-trivial optimal decay rate $d^*$.  
The optimal $r^*$ and $T^*$ balance and minimize the decay rate in $\Pch$(r,T) and $\Pdv(r,T)$. 
Hence, for Case 1, the optimal asymptotic total probability of error 
decays as follows: $$\Pt(r^*,T^*) \doteq \Pdv(r^*,T^*) \doteq \Pch(r^*,T^*) \doteq \rho^{-d^*}.$$ Note that $d^*$ is nothing but the optimal \emph{negative SNR exponent}.

In Case 2, where the source burstiness scales down slower than $\Theta(\frac{1}{\sqrt{\log\rho}})$ but faster than $\Theta(\frac{1}{\sqrt{\log\log\rho}})$, we have that $\Pt(r,T)$ is asymptotically equal to $\Pdv(r,T)$ for all $r \in (\lambda,\rmax)$ and $T\in \T$.
In this case, the decay rate of $\Pt(r,T)$ is  equal to $I(r,T)$.
In other words, the channel error (outage) probability is dominated by the delay violation probability and, hence, can be ignored.

Finally, in Case 3, when the source burstiness scales down faster than $\Theta(\frac{1}{\sqrt{\log\rho}})$, 
we have the opposite of Case 2. In Case 3, the delay violation probability is dominated by the channel error probability and, hence, can be ignored.

\subsection{Approximation of the Optimal Negative SNR Exponent}
For Case 1 in Theorem~\ref{thm:1}, we use the following approximation which is an immediate result of relaxing the integer-constrained optimizations of $I$ and $T^*$ to obtain approximated expressions with much simpler forms.
These approximations become especially useful in Section~\ref{sec:discussionOfResults}.

\begin{approximation} \label{cor:gconst}
Relaxing the integer constraints in the calculation of $I$ (as in Approximation~\ref{ApproxI}) and $T^*$ in
(\ref{main:T}) gives the following ``integer-relaxed'' approximations for
$d^*,r^*,$ and $T^*$: \begin{align} d^* &\approx d_{ir}^* := \dch(r_{ir}^*,T_{ir}^*),
\label{mainthm1:dstar}\\ T^* &\approx T^*_{ir},
\text{ and } r^* \approx r^*_{ir},\no\end{align} where, for $\delta_r$ given in (\ref{defn:deltar}) and any $T\in\T$, \beq r_{ir}^*(T) := \min\{r\in (\lambda,r_\text{max}): \dch(r,T) = \gamma\delta_r r(D\!-\!2T\!+\!1)\}, \label{mainthm1:rstarT}\eeq and
\begin{align} T_{ir}^* &= \left[\min\left\{T\in\R^+: \frac{d}{dT} \bigl(\dch(r_{ir}^*(T),T)\bigr)=0\right\}\right]_{1}^{\left\lfloor \frac{D}{2}\right\rfloor},
\label{mainthm1:Tstar}\\
r^*_{ir} &= r^*_{ir}(T^*_{ir}).\label{mainthm1:rstar}
 \end{align}
\end{approximation}

\section{Applications of the Result \label{sec:discussionOfResults}}

In this section, we apply the result of Case 1 in Theorem~\ref{thm:1} to analyze and optimize the end-to-end error probability of systems communicating delay-sensitive and bursty traffic over three outage-limited channels: SISO Rayleigh fast-fading channel, quasi-static cooperative relay channel, and quasi-static MIMO channel.

To illustrate the methodology, we restrict our attention to the case of CPE($\lambda,\mu,g,\log\rho$) arrival process where $g(\log\rho) = \log\rho$, for simplicity. Note that to better gain insights, we use the integer-relaxed approximations obtained in Approximation~\ref{cor:gconst}.

\subsection{SISO Rayleigh Fast-Fading Channel}
Our first example considers an example of SISO Rayleigh fast-fading channel, whose $\dch(r,T)=T(1-r)$ (see (\ref{eq:DMTfastFading})). Combining this with (\ref{eq:delta11}) and (\ref{mainthm1:rstarT}) gives the optimal choice of multiplexing gain when the coding duration is fixed at $T$ as \beq \ro(T) = \lambda + \frac{1-\lambda}{1+ \frac{\mu(D+1-2T)}{T}}. \label{eq:rstarT}\eeq  In addition, using (\ref{mainthm1:Tstar}), the integer-relaxed approximated optimal coding duration can be expressed as \beq \label{TstarSISO} T_{ir}^* = \left[\frac{1}{1+\frac{1}{\sqrt{2\mu}}} \ \frac{D+1}{2}\right]_1^{\left\lfloor \frac{D}{2}\right\rfloor}.\eeq Inserting $T_{ir}^*$ into (\ref{eq:rstarT}), we get the approximated optimal channel multiplexing gain as \beq \label{rstarSISO} r_{ir}^* = r^*_{ir}(T_{ir}^*) = \left[\lambda+\frac{1-\lambda}{1+\sqrt{2\mu}}\right]_{r^*_{ir}(1)}^{r^*_{ir}(\left\lfloor
\frac{D}{2}\right\rfloor)}.\eeq
Also, from (\ref{mainthm1:dstar}), the approximated optimal negative SNR exponent is given as:
\begin{align} d_{ir}^* &= T_{ir}^*(1-r_{ir}^*) \no\\ &= \left[\frac{1}{(1+\frac{1}{\sqrt{2\mu}})^2} \ \frac{(D+1)}{2} (1-\lambda)\right]_{1-r^*_{ir}(1)}^{\left\lfloor
\frac{D}{2}\right\rfloor\left(1-r^*_{ir}(\left\lfloor
\frac{D}{2}\right\rfloor)\right)}. \label{dstarSISO}
\end{align}

\begin{figure}[htbp!] \centering
\includegraphics[width=.95\linewidth]{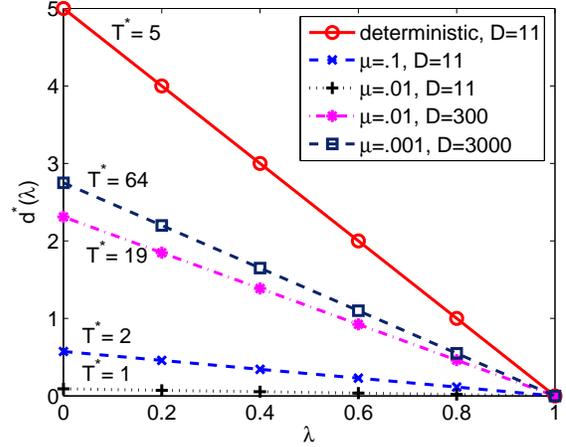} \hfill
\caption{SISO, Rayleigh
fast-fading, coherent channel. The solid line describes the DMT
($r=\lambda$).  The dashed and dotted lines describe $d_{ir}^*(\lambda)$ for various $\mu$ and $D$.
 \label{fig:FastFadingPlot}} \vspace{-0.1in}
\end{figure}

Below, we provide some observations of the above results:

$\bullet$
The above result on $d^*$ can also be interpreted as a tradeoff which describes the relation between the normalized average arrival rate,
$$\lambda := \limN \bigl(\text{average bit-arrival rate}\bigr) / N = \limN \frac{\mathbb{E}[A_t^N]}{N},$$
and the corresponding optimal negative SNR exponent
$d_{ir}^*(\lambda)$ 
as a function of the delay bound $D$, and the average packet size $1/\mu$.
For constant bit arrivals (CBR) at rate $\lambda \log \rho$, i.e., mathematically when $1/\mu \rightarrow 0$, any coding durations less than half\footnote{The first half of $D$ is spent waiting for the next coding block and the other half waiting to be decoded at the end of the block.} of $D$ (or more precisely $\left\lfloor \frac{D}{2}\right\rfloor$) and any channel multiplexing rates greater than $\lambda$ result in zero probability of delay violation. Hence, the optimal negative SNR exponent of the total error probability, denoted by $d^*_{CBR}$, is equal to the corresponding channel diversity when the optimal coding duration is at its maximum value, $\left\lfloor \frac{D}{2}\right\rfloor$,  and the channel multiplexing gain is at its minimum, $\lambda$. That is, $$d^*_{CBR}(\lambda) = \left\lfloor \frac{D}{2}\right\rfloor (1-\lambda).$$ It is not surprising that this coincides with the classical DMT. With traffic burstiness, however, the optimal negative SNR exponent $d^*_{ir}(\lambda)$ given in (\ref{dstarSISO}) is smaller than $d^*_{CBR}(\lambda)$. 
The ratio 
$$\frac{d^*_{ir}(\lambda)}{d^*_{CBR}(\lambda)} \approx \frac{1}{(1+\frac{1}{\sqrt{2\mu}})^2} \le 1$$ can be interpreted as the reduction factor on the SNR exponent in the presence of burstiness.
Figure~\ref{fig:FastFadingPlot} shows the impact of traffic burstiness (which is parameterized by $\mu$) on $d^*_{ir}(\lambda)$. 

$\bullet$ From a coding point of view, $T_{ir}^*$ is independent of the average bit-arrival rate $\lambda$. This implies that for a
fixed value of the average packet size $1/\mu$, the optimal negative SNR exponent is achieved by a fixed-duration $1\times T_{ir}^*$ code. 
Optimal codes for this setting exist for all values of $r$ and $T$ (\!\!
\cite{Petros_Thesis,TavVisUniversal_2005,Belfiore_Dynamic_AAF_journal}).
On the other hand, if $T$ is already given, the performance is optimized when the coding multiplexing gain is chosen as in (\ref{eq:rstarT}), i.e.,
$$r_{ir}^*(T) = \lambda + \frac{1-\lambda}{1+ \frac{\mu(D+1-2T)}{T}}.$$ 

$\bullet$
Since $ r_\text{max}=1$ for this SISO channel, we can verify that $r_{ir}^* \nearrow r_\text{max}$ for very bursty traffic (i.e., $1/\mu \rightarrow \infty$). That is for very bursty traffic the channel should operate close to its highest possible rate, which is the channel ergodic capacity. 


\subsubsection{Numerical Comparison of the Approximation\label{subsec:NumComparison}}

\begin{figure}[htbp!] \centering
\centering
    \subfigure[$d^*$ and $d^*_\text{ir}$ vs $D$ and $\lambda$]{
        \includegraphics[width=.8\linewidth]{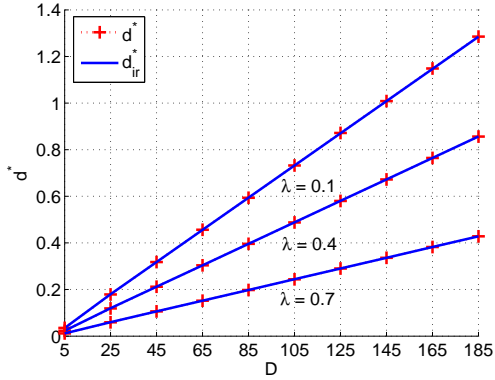}}
        \hfill
    \subfigure[$T^*$ and $T^*_\text{ir}$ vs $D$ and $\lambda$]{
        \includegraphics[width=.8\linewidth]{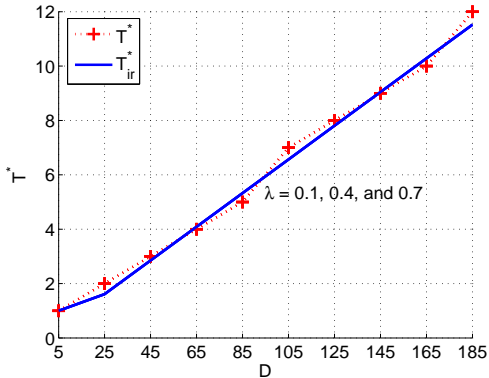}}
    \hfill
    \subfigure[$r^*$ and $r^*_\text{ir}$ vs $D$ and $\lambda$]{
        \includegraphics[width=.8\linewidth]{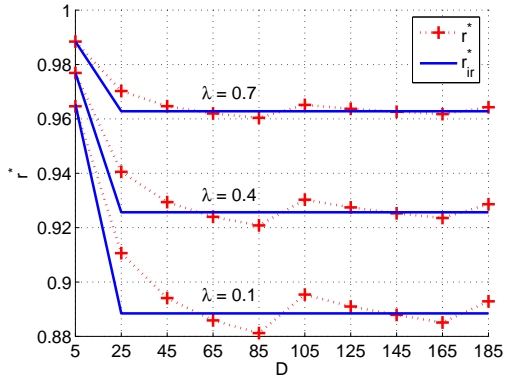}}
        \vfill
    \hfill
\hfill \caption{Comparisons of the exact values $d^*$, $T^*$, and $r^*$ (\ref{fig:comparison}a, \ref{fig:comparison}b, \ref{fig:comparison}c, respectively) and the integer-relaxed approximations $d^*_{ir}$, $T^*_{ir}$, and $r^*_{ir}$, at various $D$ and $\lambda$.
The dotted lines with markers correspond to the exact solutions while the solid
lines represent the approximated solutions.} \label{fig:comparison} \vspace{-0.2in}
\end{figure}

Before we move to the next example, we illustrate numerically that the approximations in (\ref{TstarSISO})-(\ref{dstarSISO}) well approximate their actual values in Theorem~\ref{thm:1}. 
In Figure~\ref{fig:comparison}, we show an example of a comparison at $1/\mu = 100$ and 
various values of $D$ and $\lambda$.
We observe that the approximated values match well with the exact values if $D$ is sufficiently large. The matching is remarkably good for $d^*$ and $d^*_{ir}$. Note that $r^*_{ir}$ is independent of $D$, except when $D$ is so small that $T^*_{ir} = 1$.

\subsection{Cooperative Wireless Networking with Optimal Clustering}
As studied in \cite{clusteringISIT,Allerton07}, we consider communicating bursty and delay-limited information from an information source in a
cooperative wireless relay network, shown in Figure~\ref{fig:General
snapshotCooperativeDiversity}, where the diversity benefit of user cooperation is due to encoding across space and time \cite{SenErkAaz1,LanWorTSEIEEE}.
In the absence of delay limitation, having more
cooperative users almost always improves performance.
This is not the case, though, when one considers burstiness and delay QoS requirement.
Take for example a network where the information-source node cooperates with $v$
relays, under an orthogonal amplify-and-forward (OAF) cooperative diversity scheme and half-duplex constraint. This cooperation scheme gives the DMT: $$\dch^{\text{coop}}(r) = (v+1)(1-2r).$$ Note that $r_\text{max} = 1/2$ under this protocol.
To realize this amount of diversity, the coding duration $T$ is required to be at least $2(v+1)$ channel uses or timeslots. This means that, in spite of the increase in the negative SNR exponent of the probability of decoding error with the number of cooperative relays, relaying over all nodes in the network might not be desirable as it increases the delay violations. 
Applying the result of Approximation~\ref{cor:gconst} to CPE source and the above $\dch^{\text{coop}}(r)$ with $T=2(v+1)$, the optimal performance is achieved when the nodes cooperate in clusters with $$v^* \approx v^*_{ir} = \left[\frac{D+1}{4(1+\frac{1}{\sqrt{2\mu}})} -1\right]_1^v $$ relays and transmit at multiplexing rate, $$r^* \approx r^*_{ir} = \frac{1}{2} - \frac{\frac{1}{2}-\lambda}{1+\frac{1}{\sqrt{2\mu}}}. $$ 
Note that $v^*_{ir}$ is independent of the traffic average arrival rate $\lambda$. This means that meeting the optimal tradeoff for various values of $\lambda$ does not require modifying the cluster sizes, unless the traffic burstiness (parameterized by the average packet size $1/\mu$) changes.

\begin{figure}[htbp!] \centering
\begin{center}
\begin{center}\includegraphics[width=0.99\linewidth]{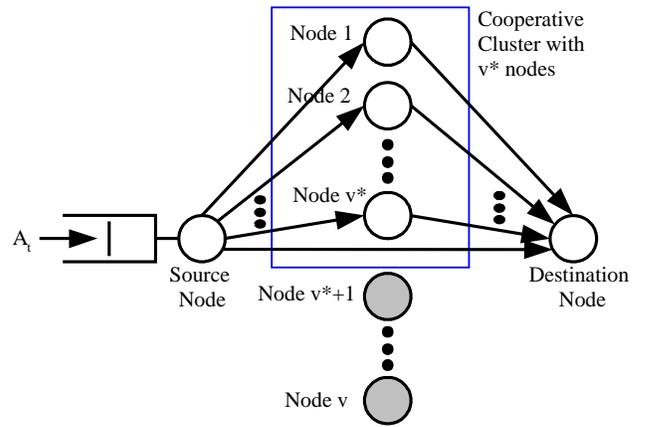} 
\end{center}
\caption{Snapshot of a wireless network, where the source node
utilizes a subset of its peers (nodes $1,2,\ldots,v^*$) as relays
for communicating with the destination. \label{fig:General
snapshotCooperativeDiversity}}
\end{center} \vspace{-0.2in}
\end{figure}

\subsection{MIMO Quasi-Static Communications}
\begin{figure}[htbp!] \centering
\includegraphics[width=.95\linewidth]{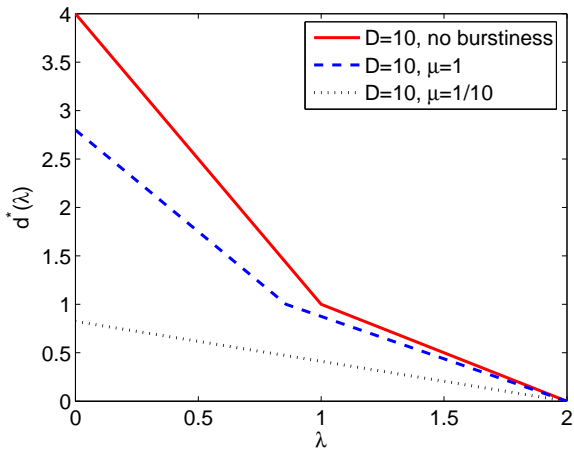} \hfill
\caption{MIMO, quasi-static, coherent 2x2 channel. $d^*$ v.s. $\lambda$ for different values of $D$ and $\mu$.    
 \label{fig:MIMOexample}} \vspace{-0.1in}
\end{figure}

In the case of the MIMO Rayleigh fading channel with $n_t$
transmit and $n_r$ receive antennas, and with complete channel
state information at the receiver (CSIR) and no CSI at the transmitter, the channel diversity gain $\dch(r)$ is shown (see \cite{ZheTse}) to be a piecewise linear function that connects
points \beq \label{dch:MIMO} (k,(n_t-k)(n_r-k)), \ \ k=0,1,\ldots,\min(n_t,n_r).\eeq  The
entire tradeoff is met if $T\ge n_t$ \cite{EliKumPawKumLu}. An
example of the effect of burstiness is shown in
Figure~\ref{fig:MIMOexample}, for the case of the $2 \times 2$
Rayleigh fading channel ($n_t=n_r=2$). By assuming that $T$ is given (not an optimizing parameter) and equal to $2$, the optimal multiplexing gain $r^*$, which balances the SNR exponents of the probabilities of delay violation and decoding error, is the solution to $\dch(r^*) = I(r^*,T=2)$. Using the approximation (\ref{approxPdvCPE}) of $I$ for CPE source, the approximation $r^*_{ir}$ is the solution to $$\dch(r^*_{ir}) - \mu(r^*_{ir}-\lambda)(D-3)=0,$$ where $\dch$ is the piecewise linear function connecting points in (\ref{dch:MIMO}). In other words, $r^*_{ir}$ is given as $$r^*_{ir} = \begin{cases} \lambda + \frac{2-\lambda}{1+\mu(D-3)}, &\text{ if } \lambda \in [1-\frac{1}{\mu(D-3)},2),\\ \lambda + \frac{4-3\lambda}{3+\mu(D-3)}, &\text{ if } \lambda \in (0,1-\frac{1}{\mu(D-3)}].\end{cases}$$
Figure~\ref{fig:MIMOexample} shows the resulting $d^*(\lambda) = \dch(r^*_{ir})$ for various values of burstiness $\mu$ and $D$.

\section{Summary and Future Work\label{sec:conclussionWiopt}}

This work offers a high-SNR asymptotic error performance analysis for communications of delay-limited and bursty information over an outage-limited channel, where errors occur either due to delay or due to erroneous decoding.  The analysis focuses on the case where there is no CSIT and no feedback, and on the static case of fixed rate and fixed length of coding blocks. This joint queue-channel analysis is performed in the asymptotic regime of high-SNR and in the assumption of smoothly scaling (with SNR) bit-arrival processes. The analysis provides closed-form expressions for the error performance, as a function of the channel and source statistics.
These expressions
identify the scaling regime of the source and channel statistics in which either delay or decoding errors are the dominant cause of errors, and the scaling regime in which a prudent choice of the coding duration and rate manages to balance and minimize these errors. That is, in this latter regime, such optimal choice manages to balance the effect of channel atypicality and burstiness atypicality.
To illustrate the results, we provide different examples that apply the results in different communication settings. We emphasize that the results hold for any coding duration and delay bound.

Many interesting extensions of the current work remain. One example is the high-SNR analysis of systems with retransmission mechanism and/or adaptive adjustment of the transmission rate and coding duration as a function of the current queue length at the transmitter. With retransmission, the diversity of the channel can be improved considerably \cite{GamCaiDamARQ} but at the cost of longer and random transmission delays.
On the other hand, we may be able to improve the system performance by adjusting the transmission rate according to the need of the bits in the queue. For example, when the queue length is short, we may reduce the transmission rate, which improves the probability of channel decoding error but possibly at the cost of longer delays of the bits that arrive later. However, since in high-SNR analysis the probability of error is asymptotically dominated by the worst case probability, it is not clear whether such adaptive transmission rate mechanism will improve the asymptotic decay rate of the probability of bit error.

In addition, this work focuses on the notion of SNR error exponent as a measure of performance. This view of communication systems provides  a tractable and intuitive characterization of various suggested schemes in the high-SNR regime. It would also be interesting to fine-tune the high-SNR asymptotic analysis presented here, for the regime of finite SNR, as well as extend it to different families of bit-arrival processes.

\appendices

\section{Proof of Proposition~\ref{GartnerEllisSmoothlyScaling}\label{append:pffact3}}

\emph{Proposition~\ref{GartnerEllisSmoothlyScaling}:}
Consider a $g$-smoothly-scaling process $A^\Ng$
with the limiting $g$-scaled log moment generation function $\Lambda$. Let $S^\Ng_t= \sum_{i=1}^t A^\Ng_i$, for $t\in \N$.
Then, for $a>\lambda t$, we have \beq
\lim_{N\rightarrow \infty} \frac{1}{g(N)} \log P\left(\frac{S^\Ng_t}{N} > a\right) = -t\Lambda^*(a/t),\eeq where $\Lambda^*$ is the convex conjugate of $\Lambda$.

\begin{proof} 
Let $n=g(N)$ and $Y^{(n)}_t = \frac{g(N)}{N} S^\Ng_t$. From (\ref{Gen:condS}) and the property of $\Lambda$ for the $g$-smoothly-scaling process, we have $$\Lambda_{Y_t}(\theta) := \lim_{n\rightarrow \infty} \frac{1}{n} \log E[e^{\theta Y^{(n)}_t}] = \Lambda_{S_t}(\theta) = t\Lambda(\theta),$$ which exists for each $\theta \in \R$ as an extended real number and is finite in a neighborhood of $\theta = 0$, essentially smooth, and lower-semicontinuous. Then, the G\"artner-Ellis theorem (Theorem 2.11 in \cite{BigQueues}) shows that $Y^{(n)}_t/n$ (which, in this case, is equivalent to $S^\Ng_t/N$) satisfies the \emph{large deviations principle} (LDP) in $\R$ with good convex rate function  $$ \Lambda_{Y_t}^*(x) := \sup_{\theta \in \R} \ \theta x - \Lambda_{Y_t}(\theta) = \sup_{\theta \in \R} \ \theta x - t \Lambda(\theta) = t\Lambda^*(x/t).$$  For $a> E[\frac{S^\Ng_t}{N}]=\lambda t$, the LDP result gives the assertion of the proposition
(see Lemma~2.6 and Theorem~2.8 in \cite{BigQueues}).
\end{proof}

\section{Proof of Results on the Asymptotic Probability of Delay Violation  \label{appen:PfPdv}}

\emph{Lemma \ref{lem:Pdv}:}
Given $g\in\G$, $T \in \T$, $r>\lambda$, a batch service of $rNT$ every $T$ timeslots, and a $g$-smoothly-scaling bit-arrival process characterized by the limiting $g$-scaled log moment generation function $\Lambda$, the decay rate of $P_\text{delay}(r,T)$ is given by the function $I$, i.e., \beq
\limN \frac{1}{g(N)} \log P_\text{delay}(r,T) = -I(r,T)\eeq
where \begin{align} &I(r,T) \no\\&= \min_{\substack{t\in\No:\\tT+T-1-k>0}} \ (tT+T-1-k) \Lambda^*\left(r+\frac{(D+1-2T)r}{tT\!+\!T\!-\!1\!-\!k}\right),\end{align}
and $k = D(\!\!\!\mod T)$. In addition, $I(r,T)$ is lower-semicontinuous and increasing on $r$.

\begin{proof} 
Let $g\in\G$, $T \in \T=\{1,2,\ldots,\left\lfloor \frac{D}{2}\right\rfloor\}$, $r>\lambda$, and $k = D(\!\!\!\mod T)$. Without loss of generality, we assume that
$I(r,T) < \infty$.

For any given SNR $\rho$ and $N=\log\rho$, there are $A^\Ng_t$ bits arriving at time $t$. The queue is being served exactly at times
$mT$, for $m\in\Z$, with an instantaneous removal of the oldest $RT=rNT$ bits. The
corresponding queue dynamics for the queue size $Q^\Ng_t$, at time
$t$, are as follows.
\beq Q^\Ng_t = \begin{cases} \left[Q^\Ng_{t-1} + A^\Ng_t - TR\right]^+, & \mbox{ if } t = mT, \ m\in\Z, \\
                    Q^\Ng_{t-1}+A^\Ng_t, & \mbox{ otherwise}, \end{cases} \label{queue dynamics} \eeq
where $Q^\Ng_{-\infty}\equiv 0$. Since the arrival process is stationary and the system started empty at time $-\infty$, then $Q^\Ng_i$ has the same steady-state distribution as that of $Q^\Ng_{mT+i}$, $m\in \Z$, for each $i = 0,\ldots,T-1$. The delay at time $i$ also has the same steady-state distribution as the delay at time $mT+i$.
Since $\Pdv(r,T)$, as a function of $r, T$, is defined as the probability of the steady-state delay being  greater than $D$, we have
\begin{align} &\Pdv(r,T) \no\\ &:=P(\text{steady-state delay of a bit} > D)\no\\
&= \frac{1}{T} \sum_{i=0}^{T-1} P(\mbox{s-s delay of a bit arriving at time } i > D), \label{eq:Pdv1} \end{align}
where the equality holds since the arrivals are independent across time. From Lemma~\ref{lem:LastBitPart2} in
Appendix~\ref{app:LastBitDefinesError}, we have that the delay violation probability of \emph{any} bit arriving at time $i$ is \emph{asymptotically} equal to the delay violation probability of the \emph{last} bit arriving at time $i$, (\ref{eq:Pdv1}) becomes \beq \Pdv(r,T) \eqg \frac{1}{T}\!\sum_{i=0}^{T-1} P(\mathcal{Q}^\Ng_i) \eqg \sum_{i=0}^{T-1} P(\mathcal{Q}^\Ng_i), \label{pf:Lemma1P1}\eeq
where $\mathcal{Q}^\Ng_i$ denotes the event that the last bit arriving at timeslot $i$ violates the delay bound $D$. This holds because $T$ is a constant independent of $\rho$.
Hence, (\ref{pf:Lemma1P1}) says that $\Pdv$ is asymptotically equal to the sum of $P(\mathcal{Q}^\Ng_i)$.

Next, we relate the event $\mathcal{Q}^\Ng_i$ to a condition on the queue length $Q^\Ng_i$, for $i=0,\ldots,T-1$.
To do this, we need to describe the condition that the delay of the last bit arriving at timeslot $i$ violates the delay bound $D$.
Upon arrival, the last bit sees $Q^\Ng_i$ bits (including itself) waiting in the queue.
Since the batch service happens exactly in multiples of $T$, the bit must wait $T-i$ timeslots for the next service to start and another $\left\lceil \frac{Q^\Ng_i}{RT}\right\rceil T$ timeslots for all $Q^\Ng_i$ bits (including the last bit) to get served and be decoded.
Hence, the last bit arriving at time $i$ violates the delay bound $D$ if, and only if, $$T-i + \left\lceil \frac{Q^\Ng_i}{RT}\right\rceil T > D.$$
Let $\Omega^\Ng$ contains all measurable random events. The condition above implies that the delay violation event for the last bit is given as \beq \mathcal{Q}^\Ng_i := \{\omega \in \Omega^\Ng: T-i + \left\lceil \frac{Q^\Ng_i(\omega)}{RT}\right\rceil T > D\}.\label{dv_cond}\eeq

Using (\ref{queue dynamics}) and (\ref{dv_cond}), we show in Lemma~\ref{lem:PdvToOverflowProb} of Appendix~\ref{lem:DominatingArrivals} that \beq \Pdv(r,T) \eqg P(\mathcal{Q}^\Ng_{T-1-k}) \eqg P(Q^\Ng_{T-1-k} > (D\!-\!T\!-\!k)R).\label{lem3:result} \eeq
Intuitively, this means that $\Pdv(r,T)$ is asymptotically equal to $P(\mathcal{Q}^\Ng_{T\!-1\!-k})$, 
equivalently $\Pdv(r,T)$ is asymptotically equal to the probability that the last bit arriving at time $T\!-1\!-k$ sees a queue length greater than $(D\!-T\!-k)R$ bits.

Finally, using (\ref{lem3:result}), what remains is to
establish that \begin{align}
&\lim_{N\rightarrow \infty} \frac{\log P(Q^\Ng_{T-k-1} >
(D\!-\!T\!-\!k)rN)}{g(N)} \no\\ &= -I(r,T) \no\\&= -\min_{\substack{t\in\No:\\tT+T-1-k>0}} \ (tT\!+\!T\!-\!k\!-\!1)
\Lambda^*\left(r+\frac{(D+1-2T)r}{tT\!+\!T\!-\!k\!-\!1}\right). \end{align}
For notational simplicity, let $i := T-1-k$ and $q :=
(D-T-k)r$. Note that $q>ri\ge 0$ since $T \in \{1,2,\ldots,\left\lfloor \frac{D}{2}\right\rfloor\}$ and $k=D(\!\!\mod T)$. 
Now, since $$\frac{q+rTt}{Tt+i} = r+\frac{(D+1-2T)r}{tT+T-k-1},$$ it is sufficient to show that \beq \limN \frac{\log P(Q^\Ng_i
> Nq)}{g(N)} = - \min_{\substack{t\in\No:\\tT+i>0}} (Tt+i) \Lambda^*(\frac{q+rTt}{Tt+i}). \label{pf:Lemma1P2}
\eeq We separately show (matching) upper and lower bounds.

First, we show the lower bound.
By using the queue dynamics in (\ref{queue dynamics}) recursively and the assumption of $Q^\Ng_{-\infty}= 0$, the queue length $Q^\Ng_i$ is related to the arrivals $A^\Ng_j$, $j\le i$, in the following manner:
\beq  Q^\Ng_i = \sup_{t\in\No} \left(\sum_{j=-tT+1}^{i} A^\Ng_j -rtTN\right), \label{eq:AfterQueueDynamics1}\eeq where we use the convention that $\sum_{j=1}^0 A^\Ng_j \equiv 0$.
Using this relation and the fact that $q>0$, we have
\begin{align*} P(Q^\Ng_{i}\!>\!Nq) &= P\left(\sup_{t\in\No} \sum_{j=-tT+1}^{i} A^\Ng_j\!-\!rtTN > Nq\right)\\
&= P\left(\sup_{\substack{t\in\No:\\tT+i>0}} \sum_{j=-tT+1}^{i} A^\Ng_j\!-\!rtTN > Nq\right). \end{align*}
Now, for any fixed $ t \in \No$ so that $tT+i>0$, we have 
\bean P(Q^\Ng_{i} > Nq) &\ge& P\left(\sum_{j=-tT+1}^{i} A^\Ng_j -rtTN > Nq\right)\\
							  &=& P\left(\sum_{j=1}^{tT+i} A^\Ng_j > N(q+rTt)\right) \\
                &=& P\left(\frac{S^\Ng_{Tt+i}}{N} > q+rTt\right).\eean
Taking the limit of both sides and using Proposition~\ref{GartnerEllisSmoothlyScaling}, 
we have
\bea \liminf_{N\rightarrow \infty} \frac{\log P(Q^\Ng_{i}
> Nq)}{g(N)} \ge -(Tt+i) \Lambda^*(\frac{q+rTt}{Tt+i}). \eea
Since $t$ is arbitrary, maximizing the RHS over $t$ gives the appropriate lower bound:
\beq \liminf_{N\rightarrow \infty} \frac{\log P(Q^\Ng_{i}
> Nq)}{g(N)} \ge -\inf_{\substack{t\in\No:\\tT+i>0}} (Tt+i) \Lambda^*(\frac{q+rTt}{Tt+i}).
\eeq

For the upper bound, we use the following result from Lemma~\ref{lem:PdvUpperBound} in Appendix~\ref{app:ProofOfLemmaPdvUpperBound}:
\beqn \limsup_{N\rightarrow \infty} \frac{\log P(Q^\Ng_i>Nq)}{g(N)}  \le -\inf_{\substack{t\in\No:\\tT+i>0}}  (Tt+i) \Lambda^*\left(\frac{q+rTt}{Tt+i}\right), \eeqn noting that the RHS is strictly greater than $-\infty$, by assumption.
Hence, the lower and upper bounds coincide and (\ref{pf:Lemma1P2}) holds.

To complete the proof, we show the properties of $I(r,T)$ for $T\in \T$. First, $I$ is increasing on $r \ge \lambda$ because $\Lambda^*(x)$ is increasing on $x \ge \lambda$ (Lemma 2.7 in \cite{BigQueues}). Second, $I(r,T)$ is lower-semicontinuous on $r$ because
 $I$ is the minimum of a number of function $\Lambda^*$ which are lower-semicontinuous (Lemma 2.7 in  \cite{BigQueues}).
\end{proof}

\emph{Approximation~\ref{ApproxI}: }
Relaxing the integer constraint in (\ref{defn:I}) gives the lower bound of $I$ as \beq I(r,T) \ge \delta_r r (D+1-2T) =: I_{ir}(r,T),\eeq where \beq \delta_r = \sup\{\theta >0: \Lambda(\theta) < \theta r\}.\label{deltarproof} \eeq 

\begin{proof} 
By the definition of $I$, we have \begin{align*} &I(r,T)\\ &= \min_{\substack{t \in\No:\\tT+T-1-k>0}} (tT\!+\!T\!-\!1\!-\!k) \Lambda^*\bigl(r+r\frac{D-2T+1}{tT\!+\!T\!-\!1\!-\!k}\bigr) \\
	&\ge \min_{\tau \in\R^+} \tau \Lambda^*\bigl(r+r\frac{D-2T+1}{\tau}\bigr) \\
	&= \delta_r r(D-2T+1), \end{align*}
where the last equality is a result of Lemma~3.4 of \cite{BigQueues} with $\delta_r$ defined as in (\ref{deltarproof}). 
\end{proof}

\section{Proof of the Main Result} \label{sect:ProofGammaConst}

\begin{proof}[Proof of Theorem~\ref{thm:1}]
Recall that: \beq \Pt(r,T) := \Pch(r,T)+(1-\Pch(r,T)) \Pdv(r,T),\label{Pt10} \eeq where, from (\ref{eq:DMT}), \beq \Pch(r,T) \doteq \rho^{-\dch(r,T)}\eeq	 and, from Lemma~\ref{lem:Pdv}, \beq \Pdv(r,T) \eqg e^{-I(r,T) g(\log\rho)}.\eeq

\underline{Case 1}: when $\lim\limits_{N\rightarrow \infty} \frac{g(N)}{N} = \gamma \in (0,\infty)$. 
We have \beq \label{eq:PdvAsDiversity} \Pdv(r,T) \doteq \rho^{-\gamma I(r,T)}\eeq and \beq \Pt(r,T) \doteq \rho^{-\min\left\{\gamma I(r,T), \ \dch(r,T)\right\}}.\label{PtotThm2} \eeq
The optimal negative SNR exponent of $\Pt$ is  \begin{align} d^* &:= \sup_{\substack{r\in (\lambda,r_\text{max})\\T \in \T}} \ \left\{\limrho -\frac{\log \Pt(r,T)}{\log\rho}\right\}\no\\
&= \sup_{\substack{r\in (\lambda,r_\text{max})\\T \in \T}} \ \ \left\{\min\left\{\gamma I(r,T), \ \dch(r,T)\right\}\right\}\no\\
&= \max_{T \in \T} \ \left\{\sup_{r\in (\lambda,r_\text{max})} \ \left\{\min\left\{\gamma I(r,T), \ \dch(r,T)\right\}\right\}\right\}. \label{dstar:proof}\end{align}
We first solve the optimization sub-problem within the bracket for any given integer $T\in\T$. Because $I(r,T)$ is increasing on $r\ge \lambda$ while $\dch(r,T)$ is strictly decreasing on $r \in [0, \rmax]$, the sub-problem is solved by the optimal choice of multiplexing gain when the coding duration is fixed at $T$ as
\beq r^*(T) := \inf\{r\in (\lambda,\rmax): \gamma I(r,T) = \dch(r,T)\}.\label{rstarT3}\eeq 
Hence, (\ref{dstar:proof}) is solved with the optimal coding duration $T^*$, given as $$T^* = \arg\max_{T\in \T} \ \gamma I(r^*(T),T),$$  
and the optimal multiplexing gain $r^*$, given as $$r^* = r^*(T^*).$$ Note that, since $I(r,T) > 0$ when $r>\lambda$ and $\dch(r,T) > 0$ when $r < r_\text{max}$, it is guaranteed that $r^*(T) \in (\lambda,r_\text{max})$.

\underline{Case 2}: when $\lim\limits_{N\rightarrow \infty} \frac{g(N)}{N} = 0$ and $\lim\limits_{N\rightarrow \infty} \frac{g(N)}{\log N} = \infty$.
In this case, for all $r\in(\lambda,\rmax)$ and all $T\in\T$, we have $\Pdv(r,T)$ asymptotically dominates $\Pch(r,T)$ and hence $\Pt(r,T)$ is asymptotically equal to $\Pdv(r,T)$. Since, for any $T\in\T$, $I(r,T)$ is increasing on $r > \lambda$, we have
\begin{align*} \sup_{\substack{r\in (\lambda,\rmax),\\T \in \T}} \ \limrho \frac{-\log \Pt(r,T)}{g(\log \rho)}
&\le \max_{T \in \T} \left\{ \sup_{r\in (\lambda,\rmax)} I(r,T)\right\}\\
&= \max_{T \in \T} \ I(\rmax,T).\end{align*}

\underline{Case 3}: when $\lim\limits_{N\rightarrow \infty} \frac{g(N)}{N} = \infty$. This case is an opposite of Case 2. Here, $\Pt(r,T)$ is asymptotically equal to $\Pch(r,T)$ for all $r \in (\lambda,\rmax)$ and all $T\in \T$. Since $\dch(r,T)$ is decreasing on $r$ and increasing on $T$, we have
\begin{align*} \sup_{\substack{r\in (\lambda,\rmax),\\T \in \T}} \ \limrho \frac{-\log \Pt(r,T)}{\log \rho} &\le \max_{T \in \T} \left\{ \sup_{r\in (\lambda,\rmax)} \dch(r,T)\right\}\\
&= \max_{T \in \T} \ \dch(\lambda,T)\\
&= \dch(\lambda,\left\lfloor \frac{D}{2}\right\rfloor).\end{align*} \vspace{-.2in}
\end{proof}

\section{Proof of Lemma ~\ref{lem:PdvToOverflowProb}} \label{lem:DominatingArrivals}
In this appendix, we prove the following lemma which is used in Appendix~\ref{appen:PfPdv}.

\begin{lem} \label{lem:PdvToOverflowProb}
Consider $g\in\G$, $T \in \T=\{1,\ldots,\left\lfloor \frac{D}{2}\right\rfloor\}$, $r>\lambda$, a family of  $g$-smoothly-scaling bit-arrival processes characterized by the limiting $g$-scaled log moment generation function $\Lambda$, and a periodic batch service of $rNT$ bits at timeslots $mT$, $m\in\Z$. 
Let $Q^\Ng_i$ be the queue length at time $i \in\{0,\ldots,T-1\}$. Then, the event $\mathcal{Q}^\Ng_{T-k-1}$, defined as $$\mathcal{Q}^\Ng_{T-k-1} = \left\{\omega \in \Omega^\Ng:  k+1 + \left\lceil \frac{Q^\Ng_{T-k-1}(\omega)}{RT}\right\rceil T > D \right\},$$ with $k=D (\!\!\!\mod T)$,
asymptotically dominates $\Pdv(r,T)$. In other words,
\beq \Pdv(r,T) \eqg \Pr\bigl(Q^\Ng_{T-k-1}
> (D-T-k)r \log\rho\bigr). \label{result:lem:PdvToOverflowProb} \eeq
\end{lem}

\begin{proof}
Let $k=D (\!\!\!\mod T)$ and $i \in \{0,\ldots,T-1\}$.
Recall from (\ref{dv_cond}) that $$\mathcal{Q}^\Ng_i = \{\omega \in \Omega^\Ng: T-i + \left\lceil \frac{Q^\Ng_i(\omega)}{RT}\right\rceil T > D\}.$$ Now using the observation that, for any $x,y \in\R$, $$\left\lceil x\right\rceil > y \Leftrightarrow \left\lceil x\right\rceil > \left\lfloor y\right\rfloor \Leftrightarrow x > \left\lfloor y\right\rfloor,$$ we have
\begin{align} \mathcal{Q}^\Ng_i	&= \left\{\omega: \frac{Q^\Ng_i(\omega)}{RT} > \left\lfloor \frac{D+i-T}{T}\right\rfloor\right\}\no\\
	&= \begin{cases} \hspace{-2pt} \bigl\{ \omega  :  Q^\Ng_i(\omega) > (D\!-\!T\!-\!k)R\bigr\}, & \hspace{-2pt}  i\in[0,T\!-\!k\!-\!1]\\
\hspace{-2pt}\bigl\{ \omega  :  Q^\Ng_i(\omega) > (D\!-\!k)R\bigr\}, &
\hspace{-2pt} i\in[T\!-\!k,T\!-\!1].\end{cases} \label{eq:casesForQi}
	\end{align}

On the other hand, (\ref{pf:Lemma1P1}) implies that
\begin{align} &\Pdv(r,T) \no\\
&\eqg \sum_{i=0}^{T-1} P(\mathcal{Q}^\Ng_i)\no\\
&= \sum_{i=0}^{T-k-1} P(\mathcal{Q}^\Ng_i) + \sum_{i=T-k}^{T-1} P(\mathcal{Q}^\Ng_i)  \no\\
&\underset{(a)}{\eqg} P(\mathcal{Q}^\Ng_{T-k-1}) + P(\mathcal{Q}^\Ng_{T-1})\no\\
&\eqg \max\{P(\mathcal{Q}^\Ng_{T-k-1}),P(\mathcal{Q}^\Ng_{T-1})\}\no\\
&\underset{(b)}{=} \max\{P(Q^\Ng_{T-k-1}\!>\!(D\!-\!T\!-\!k)R),P(Q^\Ng_{T-1}\!>\!(D\!-\!k)R)\} \no\\
&\underset{(c)}{=} P(Q^\Ng_{T-k-1}\!>\!(D\!-\!T\!-\!k)R),
\label{max1} \end{align}
where the equality in (b) is from (\ref{eq:casesForQi}). Next, we establish the (asymptotic) equalities (a) and (c). For (a), we first need to show that
\beq \sum_{j=0}^{T-k-1} P(\mathcal{Q}^\Ng_j) \eqg
P(\mathcal{Q}^\Ng_{T-k-1}).\label{apart1} \eeq
To establish this, we first observe that \beq Q^\Ng_j(\omega) = Q^\Ng_i(\omega) +
\underbrace{A^\Ng_{i+1}(\omega) + \ldots + A^\Ng_{j}(\omega)}_{\geq 0} \ge
Q^\Ng_i(\omega), \label{queuedynamic2} \eeq for all
$\omega\in \Omega^\Ng$ and 
$0\le i \le j \le T-1$.
Hence, from (\ref{eq:casesForQi}), we have
$$P(\mathcal{Q}^\Ng_{T-k-1})\ge P(\mathcal{Q}^\Ng_i), \ \ i\in\{0,\ldots,T-k-1\},$$ which implies
\beq \sum_{i=0}^{T-k-1} P(\mathcal{Q}^\Ng_i) \le (T-k)P(\mathcal{Q}^\Ng_{T-k-1}).\label{sumP1}\eeq On the other hand, from the non-negativity of probability, we have
\beq \sum_{i=0}^{T-k-1} P(\mathcal{Q}^\Ng_i) \ge P(\mathcal{Q}^\Ng_{T-k-1}).\label{sumP2}\eeq  Combining (\ref{sumP1}) and (\ref{sumP2}), we have (\ref{apart1}). Similarly, we can show that
\beq \sum_{j=T-k}^{T-1} P(\mathcal{Q}^\Ng_j) \eqg P(\mathcal{Q}^\Ng_{T-1}).\label{sumP3}\eeq Combining (\ref{apart1}) and (\ref{sumP3}), equality (a) in (\ref{max1}) is established.

To establish equality (c), it is sufficient to show that
\beq P(Q^\Ng_0\!>\! D'R) \le P(Q^\Ng_j\!>\!D'R) \le P(Q^\Ng_0> (D'\!-\!T)R), \label{ineq:PQi} \eeq for any $D'>T$ and $j\in\left\{0,\ldots,T-1\right\}$.
This is because for $j_1=T-1$ and $D'_1=D-k$,
we get $$ P\bigl(Q^\Ng_{T-1}>(D-k)R\bigr) \le P\bigl(Q^\Ng_{0}>(D-T-k)R\bigr),$$ while for $j_2=T\!-k\!-1$ and $D'_2=D\!-T\!-k$, we get
$$P\bigl(Q^\Ng_{0}>(D-T-k)R\bigr) \le P\bigl(Q^\Ng_{T-k-1}>(D-T-k)R\bigr),$$ asserting (c).

We  prove (\ref{ineq:PQi}) in two steps.
The lower bound directly follows from (\ref{queuedynamic2}), i.e., $$Q^\Ng_j(\omega) \ge Q^\Ng_0(\omega), \ \forall \omega \in \Omega^\Ng. $$
For the upper bound, we notice that, for $D'>T$ and \begin{align*} \omega &\in \left\{\omega \in \Omega^\Ng: Q^\Ng_j(\omega) > D'R\right\}\\ & \subseteq \left\{\omega \in \Omega^\Ng: Q^\Ng_j(\omega) > TR\right\},\end{align*} 
$Q^\Ng_{j}(\omega)$ is related to $Q^\Ng_{T}(\omega)$ as 
\bean Q^\Ng_{T}(\omega) &=& [Q^\Ng_{j}(\omega) +
A^\Ng_{j+1}(\omega)+\cdots+A^\Ng_{T}(\omega) - TR]^+\\
	&=& Q^\Ng_{j}(\omega) + A^\Ng_{j+1}(\omega)+\cdots+A^\Ng_{T}(\omega) - TR, \eean where $[\cdot]^+$ is removed. 
As a result, we have \begin{align*} &P(Q^\Ng_j > D'R)\\
&=P(Q^\Ng_{T} - \left\{A^\Ng_{j+1}+\ldots+A^\Ng_{T}\right\} + TR  > D'R)\\
&\le P(Q^\Ng_{T} > (D'-T) R) \\ &= P(Q^\Ng_0 > (D'-T)R), \end{align*} where the
last equality holds since $Q^\Ng_T$ and $Q^\Ng_0$ have the same stationary distribution.
\end{proof}

\section{Proof of Lemma~\ref{lem:LastBitPart2}} \label{app:LastBitDefinesError}
This appendix shows that the average probability of delay violation for bits
that arrive at time $i$ is asymptotically equal to the corresponding
probability for the last bit arriving at that time. The proof is mainly based on the definition of the  $g$-smoothly-scaling process.

\begin{lem} \label{lem:LastBitPart2}
Consider $g\in\G$ and a family of  $g$-smoothly-scaling bit-arrival processes $((A_t^\Ng,t\in\Z),N\in\N)$, characterized by the limiting $g$-scaled log moment generation function $\Lambda$.
For any given $N$, 
let $W^\Ng$ be a random
variable having the same distribution as the steady-state
distribution of the delay of a randomly chosen bit that arrives at
time $i\in \{0,\ldots,T-1\}$ while $Z^\Ng$ is a random variable having a
distribution that is identical to the steady-state distribution of
the delay for the last bit that arrives during time $i$. Then, for
any $D>0$, \beq P(W^\Ng>D) \eqg P(Z^\Ng>D). \label{lem:LastBitPart2:main} \eeq
\end{lem}

\begin{proof}
We show (\ref{lem:LastBitPart2:main}) by showing the upper bound: \beq P(W^\Ng > D) \le P(Z^\Ng > D) \label{lem4:upperbd} \eeq and the lower bound:
\beq  P(W^\Ng > D) \geg P(Z^\Ng > D). \label{lem4:lowerbd}
\eeq
The upper bound is an immediate consequence of $W^\Ng(\omega) \le Z^\Ng(\omega)$ for $\omega \in \Omega^\Ng$. Below we prove the lower bound. We have 
\beq P(W^\Ng > D) =  \sum_{a \in \N} P(W^\Ng > D |A^\Ng_i = a)P(A^\Ng_i=a).\label{PW1}\eeq 
Now, given that $A^\Ng_i = a$ bits arrive at time $i$, we index the
$a$ bits as bit 1 to $a$, where bit 1 arrives first and bit $a$
arrives last.  Given $A^\Ng_i = a$, we let $W^\Ng_j$ to be the steady-state
delay of the $j$-th bit, $j \in \left\{1,\ldots,a\right\}$. Since the bit can have any index, from $1$ to $a$, with equal probability of $1/a$, we have \beqn P(W^\Ng\!>\!D | A^\Ng_i\!=\!a) = \frac{1}{a} \sum_{j=1}^a P(W^\Ng_j\!>\!D| A^\Ng_i\!=\!a).\eeqn Ignoring all but the last term in the sum, we have \bean P(W^\Ng> D | A^\Ng_i\!=\!a) &\ge& \frac{1}{a} P(W^\Ng_a>D|A^\Ng_i\!=\!a)\\ &=& \frac{1}{a} P(Z^\Ng > D| A^\Ng_i\!=\!a),\eean where the equality is a result of how $Z^\Ng$ is defined.
This means that 
\bean P(W^\Ng\!>\!D) &\ge&  \sum_{a \in \N} \frac{1}{a}P(Z^\Ng\!>\!D |A^\Ng_i\!=\!a)P(A^\Ng_i\!=\!a)\\ &=& \sum_{a \in \N} \frac{1}{a}P(Z^\Ng\!>\!D \text{ and } A^\Ng_i\!=\!a).\eean
Now, for a given $\beta > 0$, define $$B^\Ng := \{b\in\N: b < e^{\beta g(N)}\}.$$
We can further lower bound $P(W^\Ng\!>\!D)$ as follows:
\begin{align} &P(W^\Ng > D)\no\\
	&\ge \sum_{a\in B^\Ng} \frac{1}{a} P(Z^\Ng > D \text{ and } A^\Ng_i = a)\no\\
	&\ge e^{-\beta g(N)} \sum_{a\in B^\Ng} P(Z^\Ng > D \text{ and } A^\Ng_i = a)\no\\
	&= e^{-\beta g(N)} P(Z^\Ng > D \text{ and } A^\Ng_i \in B^\Ng),\label{lowerboundPWD}
\end{align} where the second inequality holds because $1/a > e^{-\beta g(N)}$ for any $a \in B^\Ng$.

Next, we show that $P(A^\Ng_i \in B^\Ng) \rightarrow 1$ as
$N\rightarrow \infty$. 
We do this by using the definition of the
$g$-smoothly-scaling process: there exists $\theta >0$ such that $$\limN \frac{\log
E[e^{\theta A^\Ng_i g(N)/N}]}{g(N)} = \Lambda(\theta) < \infty.$$ Hence, for any $\epsilon > 0$, there exists $N_0=N_0(\epsilon)$ such that for all $N>N_0$, we have \beq g(N)(\Lambda(\theta)+\epsilon) > \log E[e^{\theta A^\Ng_i g(N)/N}].\label{lem4eq1} \eeq The RHS can be lower-bounded, for any $a_1 \in \N$:
\begin{align*} \log E[e^{\theta A^\Ng_i g(N)/N}]
&= \log \left(\sum_{a\in \N} P(A^\Ng_i = a) e^{\theta a g(N)/N} \right)\\
&\ge \log \left(\sum_{a \ge a_1} P(A^\Ng_i = a) e^{\theta a g(N)/N} \right)\\
&\ge \log \left(P(A^\Ng_i \ge a_1) e^{\theta a_1 g(N)/N} \right) \\
&= \theta a_1 \frac{g(N)}{N} + \log  P(A^\Ng_i \ge a_1). \end{align*}
This together with (\ref{lem4eq1}) gives \beqn \log P(A^\Ng_i \ge a_1) < g(N)[\Lambda(\theta)+\epsilon - \frac{\theta a_1}{N}],\eeqn for all $a_1\in \N$.
Now, we select $a_1 = e^{\beta g(N)}$ to get \begin{align*} \log\left(1-P(A^\Ng_i \in B^\Ng)\right) &=\log P(A^\Ng_i \ge e^{\beta g(N)})\\ &< g(N)[\Lambda(\theta)+\epsilon - \frac{\theta e^{\beta g(N)}}{N}]. 
\end{align*} 
Since $\lim\limits_{N\rightarrow \infty} \frac{g(N)}{\log N} = \infty$, we, then, have \beq P\left(A^\Ng_i \in B^\Ng\right) \rightarrow 1.\label{PA1}\eeq Finally, combining (\ref{PA1}) and (\ref{lowerboundPWD}) implies that, for any $\beta>0$, $$\limN \frac{\log P(W^\Ng > D)}{g(N)} \ge \limN \frac{\log P(Z^\Ng\!>\!D)}{g(N)} -\beta.$$ Since $\beta$ can be chosen arbitrarily small, we have the lower bound in (\ref{lem4:lowerbd}), hence the assertion of the lemma.
\end{proof}

\section{Proof of Lemma~\ref{lem:PdvUpperBound}\label{app:ProofOfLemmaPdvUpperBound}}
In this appendix, we prove the following lemma which is used in Appendix~\ref{appen:PfPdv}.

\begin{lem} \label{lem:PdvUpperBound}
Consider $g\in\G$, $T \in \T=\{1,\ldots,\left\lfloor \frac{D}{2}\right\rfloor\}$, $r>\lambda$, a family of  $g$-smoothly-scaling bit-arrival processes characterized by the limiting $g$-scaled log moment generation function $\Lambda$, and a periodic batch service of $rNT$ bits at timeslots $mT$, $m\in\Z$. Let $Q^\Ng_i$ be the queue length at time $i \in\{0,\ldots,T-1\}$. Then, for $q>ir$, we have %
\beq \limsup_{N\rightarrow \infty} \frac{\log P(Q^\Ng_i\!>\!Nq)}{g(N)}  \le -\inf_{\substack{t\in\No:\\tT+i>0}} (Tt+i) \Lambda^*\left(\frac{q+rTt}{Tt+i}\right), \label{Lem5:statement} \eeq assuming that the RHS is strictly greater than $-\infty$.
\end{lem}

\begin{proof}
The proof uses the same technique as in \cite[Lemma 1.10 and 1.11]{BigQueues}. 
Using (\ref{eq:AfterQueueDynamics1}), we have the following bound:
\begin{align} P(Q^\Ng_{i}\!>\!Nq) &= P\left(\sup_{\substack{t\in\No:\\tT+i>0}} \sum_{j=-tT+1}^{i} A_j^\Ng\!-\!rtTN\!>\!Nq\right) \no\\
        &= P\left(\sup_{\substack{t\in\No:\\tT+i>0}} S^\Ng_{tT+i} - rtTN > Nq\right) \no\\
        &\leq \sum_{t:t>-\frac{i}{T}} P\left(S^\Ng_{tT+i} > N(q+rTt)\right). \no\end{align}
Now, for any fixed $t_0 \in \N$, we have
\begin{align} P(Q^\Ng_{i}\!>\!Nq)&\le \sum_{-\frac{i}{T}<t\le t_0} P(S^\Ng_{tT+i} > N(q+rTt))\no\\
        & \ \ \ \ + \sum_{t>t_0} P(S^\Ng_{tT+i} > N(q+rTt)). \label{eq:LowerBoundPart}\end{align}
Employing the principle of the largest term\footnote{The principle of the largest term \cite[Lemma 2.1]{BigQueues}: Let $a_n$ and $b_n$ be sequences  in $\R^+$. If $n^{-1}\log a_n \rightarrow a$ and $n^{-1}\log b_n \rightarrow b$, then $n^{-1}\log (a_n+b_n) \rightarrow \max(a,b)$. This extends easily to finite sums.} gives
\begin{align} & \limsup_{N\rightarrow \infty} \frac{\log P(Q^\Ng_{i} > Nq)}{g(N)}  \no\\
&\leq \max\biggl(\max_{-\frac{i}{T}<t\le t_0} \limsup_{N\rightarrow \infty} \frac{\log P(S^\Ng_{tT+i} > N(q+rTt))}{g(N)},  \no \\
& \hspace{0.5cm} \limsup_{N\rightarrow \infty} \frac{1}{g(N)} \log \sum_{t>t_0}  P(S^\Ng_{tT+i} >
  N(q+rTt)) \biggr).\label{lem4:eq1} \end{align}

For the first term (the $t\le t_0$ term) in the maximum, we use Proposition~\ref{GartnerEllisSmoothlyScaling} to get
\begin{align} &\max_{-\frac{i}{T}<t\le t_0} \limsup_{N\rightarrow \infty} \frac{1}{g(N)} \log P\left(\frac{S^\Ng_{tT+i}}{N} > q+rTt\right) \no\\
  &\leq \max_{-\frac{i}{T}<t\le t_0} -(Tt+i)\Lambda^*\left(\frac{q+rTt}{Tt+i}\right)\no\\
  &\le - \inf_{\substack{t\in\No:\\tT+i>0}} \ (Tt+i)\Lambda^*\left(\frac{q+rTt}{Tt+i}\right),
\label{eq:FiniteAndInfititeParts2} \end{align} which is the RHS of (\ref{Lem5:statement}) and finite by assumption.

Now, we show that we can select $t_0$ appropriately such that the second term (the $t > t_0$ term) in the RHS of (\ref{lem4:eq1}) is also no greater than the RHS of (\ref{Lem5:statement}).
In other words, we show that there exists $t_0$ such that \begin{align} &\limsup_{N\rightarrow \infty} \frac{1}{g(N)}\log \sum_{t>t_0}  P\left(S^\Ng_{tT+i} >
 N(q+rTt)\right)\no\\ &\le - \inf_{\substack{t\in\No:\\tT+i>0}} \ (Tt+i)\Lambda^*\left(\frac{q+rTt}{Tt+i}\right). \label{eq:infinite_0}\end{align}
This is shown by proving that there exist some $\theta >0$ and $\epsilon >0$ such that \begin{align} &\limsup_{N\rightarrow \infty} \frac{1}{g(N)} \log \sum_{t>t_0} P\left(S^\Ng_{tT+i} > N(q+rTt)\right)\no\\ &\le -\ep \theta  \bigl((t_0+1)T+i\bigr),\label{boundinfinite} \end{align} for all $t_0 \in \N$.   
Now, selecting $$t_0 =  \left\lceil \frac{1}{\ep\theta  T} \inf_{\substack{t\in\No:\\tT+i>0}} \ (Tt+i)\Lambda^*\left(\frac{q+rTt}{Tt+i}\right)\right\rceil$$  provides (\ref{eq:infinite_0}).

To prove (\ref{boundinfinite}), we first use Chernoff bound as follows:
\begin{align} &\sum_{t>t_0}  P(S^\Ng_{tT+i} > N(q+rTt)) \no\\
	&= \sum_{t>t_0} P\left(e^{\frac{\theta g(N)}{N} S^\Ng_{tT+i}} > e^{\frac{\theta g(N)}{N} N(q+rTt)}\right)\no\\
	&\le \sum_{t>t_0}  e^{-\theta g(N) (q+rTt)} E[e^{\theta S^\Ng_{tT+i} \frac{g(N)}{N}}] \no\\
	&=\sum_{t>t_0} e^{-\theta g(N) (q+rTt)} (E[e^{\theta A^\Ng_1 \frac{g(N)}{N}}])^{tT+i} \no\\
	&= \sum_{t>t_0} \exp(-g(N) (tT+i) \times \no\\& \hspace{.8in} \left[\theta\left(\frac{q+rtT}{tT+i}\right) -
\frac{\log E[e^{\frac{\theta g(N)}{N} A^\Ng_1}]}{g(N)}\right]),
\label{eq:infinitePartOfSum} \end{align} where $\theta$ is an arbitrary positive scalar and the second equality is a consequence of i.i.d. assumption on $A^\Ng_t$.

Next, we use the convexity of $\Lambda$ and the fact that $\Lambda'(0) = \lambda < r$ (Remark~\ref{rem:Lambda}) to establish that there exist some $\theta >0$ and $\epsilon >0$ for which \beq \Lambda(\theta) < \theta (r-2\epsilon).\label{Lamdabound}\eeq

On the other hand, from (\ref{Gen:condA}), we know that $\frac{\log E[e^{\frac{\theta g(N)}{N} A^\Ng_1}]}{g(N)} \rightarrow \Lambda(\theta)$. This means that there exists a  $N_0=N_0(\theta,\epsilon)$ such that, for all $N>N_0$, $$\frac{\log E[e^{\frac{\theta g(N)}{N} A^\Ng_1}]}{g(N)} < \Lambda(\theta) + \theta \epsilon.$$ Combining this with (\ref{Lamdabound}), we have \beq  \frac{\log E[e^{\frac{\theta g(N)}{N} A^\Ng_1}]}{g(N)} < \theta (r-2\ep) + \theta \ep = \theta(r-\ep),\label{Lamdabound2}\eeq  for all $N>N_0$.

Hence, using (\ref{Lamdabound2}), the term inside the square bracket in (\ref{eq:infinitePartOfSum}) can be bounded, uniformly over all $t>t_0$, as
\begin{align} &\theta\left(\frac{q+rtT}{tT+i}\right) -
\frac{\log E[e^{\frac{\theta g(N)}{N} A^\Ng_1}]}{g(N)}\no\\ &=  \theta\left(r + \frac{q-ir}{tT+i}\right) -
\frac{\log E[e^{\frac{\theta g(N)}{N} A^\Ng_1}]}{g(N)}\no\\
&> \theta r - \frac{\log E[e^{\frac{\theta g(N)}{N} A^\Ng_1}]}{g(N)}\no\\
&> \theta r - \theta(r-\ep)\no\\
&= \theta \ep,\label{Lamdabound3}
\end{align} where the first equality holds because $q>ir$, by assumption. 

Inserting (\ref{Lamdabound3}) into (\ref{eq:infinitePartOfSum}), we have (\ref{boundinfinite}):
 \begin{align*} &\limsup_{N\rightarrow \infty} \frac{1}{g(N)} \log \sum_{t>t_0} P\left(S^\Ng_{tT+i} > N(q+rTt)\right)\\ &\le \limsup_{N\rightarrow \infty} \frac{1}{g(N)} \log \sum_{t>t_0} \exp\left(-g(N) (tT+i) \theta \ep\right)\no\\
 &= \limsup_{N\rightarrow \infty} \frac{1}{g(N)} \log\left(\frac{e^{-g(N) \theta \ep ((t_0+1)T+i)}}{1-e^{-g(N)\theta
\ep T}}\right) \no\\
 &= -\ep \theta  \bigl((t_0+1)T+i\bigr), \end{align*} and, hence, the assertion of the lemma.
\end{proof}



\begin{thebibliography}{10}
\providecommand{\url}[1]{#1}
\csname url@samestyle\endcsname
\providecommand{\newblock}{\relax}
\providecommand{\bibinfo}[2]{#2}
\providecommand{\BIBentrySTDinterwordspacing}{\spaceskip=0pt\relax}
\providecommand{\BIBentryALTinterwordstretchfactor}{4}
\providecommand{\BIBentryALTinterwordspacing}{\spaceskip=\fontdimen2\font plus
\BIBentryALTinterwordstretchfactor\fontdimen3\font minus
  \fontdimen4\font\relax}
\providecommand{\BIBforeignlanguage}[2]{{%
\expandafter\ifx\csname l@#1\endcsname\relax
\typeout{** WARNING: IEEEtran.bst: No hyphenation pattern has been}%
\typeout{** loaded for the language `#1'. Using the pattern for}%
\typeout{** the default language instead.}%
\else
\language=\csname l@#1\endcsname
\fi
#2}}
\providecommand{\BIBdecl}{\relax}
\BIBdecl

\bibitem{ZheTse}
L.~Zheng and D.~Tse, ``Diversity--multiplexing: a fundamental tradeoff in
  multiple--antenna channels,'' \emph{{IEEE} Trans. Inf. Theory}, vol.~49,
  no.~5, pp. 1073--1096, May 2003.

\bibitem{OzarowShamaiWyner1994}
L.~Ozarow, S.~Shamai, and A.~Wyner, ``Information theoretic considerations for
  cellular mobile radio,'' \emph{{IEEE} Trans. Veh. Technol.}, vol.~43, no.~2,
  pp. 359--378, 1994.

\bibitem{BerryLargeDelay}
R.~Berry and R.~Gallager, ``Communication over fading channels with delay
  constraints,'' \emph{{IEEE} Trans. Inf. Theory}, vol.~48, no.~5, pp.
  1135--1149, 2002.

\bibitem{BerrySmallDelayITA06}
R.~Berry, ``Optimal power--delay trade-offs in fading channels: small delay
  asymptotics,'' in \emph{Information Theory and Applications - Inaugural
  workshop}, San Diego, CA, Feb. 2006.

\bibitem{RajanSabharwalAazhang04}
D.~Rajan, A.~Sabharwal, and B.~Aazhang, ``Delay--bounded packet scheduling of
  bursty traffic over wireless channels,'' \emph{{IEEE} Trans. Inf. Theory},
  vol.~50, no.~1, pp. 125--144, 2004.

\bibitem{Negi}
R.~Negi and S.~Goel, ``An information--theoretic approach to queuing in wireless
  channels with large delay bounds,'' in \emph{IEEE Global Telecommunications
  Conference ( GLOBECOM '04)}, vol.~1, 2004, pp. 116--122 Vol.1.

\bibitem{BetteshShamai}
I.~Bettesh and S.~Shamai, ``Optimal power and rate control for minimal average
  delay: The single--user case,'' \emph{{IEEE} Trans. Inf. Theory}, vol.~52,
  no.~9, pp. 4115--4141, 2006.

\bibitem{LiuParag}
L.~Liu, P.~Parag, J.~Tang, W.-Y. Chen, and J.-F. Chamberland, ``Resource
  allocation and quality of service evaluation for wireless communication
  systems using fluid models,'' \emph{{IEEE} Trans. Inf. Theory}, vol.~53,
  no.~5, pp. 1767--1777, 2007.

\bibitem{EffectiveCapac}
D.~Wu and R.~Negi, ``Effective capacity: a wireless link model for support of
  quality of service,'' \emph{{IEEE} Trans. Wireless Commun.}, vol.~2, no.~4,
  pp. 630--643, Jul. 2003.

\bibitem{Dembo}
A.~Dembo and O.~Zeitouni, \emph{Large {D}eviations techniques and
  applications}, 2nd~ed.\hskip 1em plus 0.5em minus 0.4em\relax Springer, 1998.

\bibitem{Weiss86}
A.~Weiss, ``A new technique for analyzing large traffic systems,''
  \emph{Advances in Applied Probability}, vol.~18, pp. 506--532, 1986.

\bibitem{BotDuf95}
D.~D. Botvich and N.~G. Duffield, ``Large deviations, the shape of the loss
  curve, and economies of scale in large multiplexers,'' \emph{Queueing
  System}, vol.~20, pp. 293--320, 1995.

\bibitem{CouWeb96}
C.~Courcoubetis and R.~Weber, ``Buffer overflow asymptotics for a buffer
  handling many traffic sources,'' \emph{Journal of Applied Probability},
  vol.~33, pp. 886--903, 1996.

\bibitem{BigQueues}
A.~Ganesh, N.~O'Connell, and D.~Wischik, \emph{Big Queues}.\hskip 1em plus
  0.5em minus 0.4em\relax Springer--Verlag, 2004.

\bibitem{MIMOMac}
S.~Kittipiyakul and T.~Javidi, ``Optimal operating point for {MIMO} multiple
  access channel with bursty traffic,'' \emph{{IEEE} Trans. Wireless Commun.},
  vol.~6, no.~12, Dec. 2007.

\bibitem{Somsak}
------, ``Optimal operating point in {MIMO} channel for delay--sensitive and
  bursty traffic,'' in \emph{IEEE Int. Symp. Information Theory}, Seattle,
  Washington, USA, Jul. 2006.

\bibitem{EliKitJav_WiOpt_2007}
P.~Elia, S.~Kittipiyakul, and T.~Javidi, ``On the
  {R}esponsiveness--{D}iversity--{M}ultiplexing tradeoff,'' in \emph{5th Intl.
  Symp. on Modeling and Optimization in Mobile, Ad Hoc, and Wireless Networks},
  Apr. 2007.

\bibitem{Goldsmith}
T.~Holliday and A.~Goldsmith, ``Joint source and channel coding for {MIMO}
  systems,'' in \emph{Allerton Conf. on Comm., Control, and Computing}, 2004.

\bibitem{EliKumPawKumLu}
P.~Elia, K.~Kumar, S.~Pawar, P.~Kumar, and H.-F. Lu, ``Explicit, minimum--delay
  space--time codes achieving the diversity--multiplexing gain tradeoff,''
  \emph{{IEEE} Trans. Inf. Theory}, vol.~52, no.~9, pp. 3869--3884, 2006.

\bibitem{Belfiore_Dynamic_AAF_journal}
S.~Yang and J.-C. Belfiore, ``Optimal {S}pace–{T}ime codes for the {MIMO}
  amplify--and--forward cooperative channel,'' \emph{{IEEE} Trans. Inf. Theory},
  vol.~53, no.~2, pp. 647--663, 2007.

\bibitem{TavVisUniversal_2005}
S.~Tavildar and P.~Viswanath, ``Approximately universal codes over slow--fading
  channels,'' \emph{{IEEE} Trans. Inf. Theory}, vol.~52, no.~7, pp. 3233--3258,
  2006.

\bibitem{GamCaiDam}
H.~El~Gamal, G.~Caire, and M.~Damen, ``Lattice coding and decoding achieve the
  optimal diversity--multiplexing tradeoff of {MIMO} channels,'' \emph{{IEEE}
  Trans. Inf. Theory}, vol.~50, no.~6, pp. 968--985, 2004.

\bibitem{CaiEliRaj}
P.~Elia, G.~Caire and K.~R. Kumar, ``Space--time coding: an overview,''
  \emph{Journal of Communications Software and Systems}, Oct. 2006.

\bibitem{EliSetKum}
P.~Elia, B.~Sethuraman, and P.~Vijay~Kumar, ``Perfect space–-time codes for any
  number of antennas,'' \emph{{IEEE} Trans. Inf. Theory}, vol.~53, no.~11, pp.
  3853--3868, 2007.

\bibitem{TseVisZhe_MultAccess_2004}
D.~Tse, P.~Viswanath, and L.~Zheng, ``Diversity--multiplexing tradeoff in
  multiple--access channels,'' \emph{{IEEE} Trans. Inf. Theory}, vol.~50, no.~9,
  pp. 1859--1874, Sep. 2004.

\bibitem{LanWorTSEIEEE}
J.~Laneman, D.~Tse, and G.~Wornell, ``Cooperative diversity in wireless
  networks: Efficient protocols and outage behavior,'' \emph{{IEEE} Trans. Inf.
  Theory}, vol.~50, no.~12, pp. 3062--3080, 2004.

\bibitem{AzaGamSch}
K.~Azarian, H.~El~Gamal, and P.~Schniter, ``On the achievable
  diversity--multiplexing tradeoff in half--duplex cooperative channels,''
  \emph{{IEEE} Trans. Inf. Theory}, vol.~51, no.~12, pp. 4152--4172, 2005.

\bibitem{Petros_Thesis}
P.~Elia, ``Asymptotic universal optimality in wireless multi--antenna
  communications and wireless networks,'' Ph.D. dissertation, USC, 2006.

\bibitem{Viniotis98}
Y.~Viniotis, \emph{Probability and Random Processes for Electrical
  Engineers}.\hskip 1em plus 0.5em minus 0.4em\relax McGraw-Hill, 1998.

\bibitem{clusteringISIT}
P.~Elia, S.~Kittipiyakul, and T.~Javidi, ``Cooperative diversity in wireless
  networks with stochastic and bursty traffic,'' in \emph{IEEE Int. Symp.
  Information Theory}, Nice, France, Jun. 2007.

\bibitem{Allerton07}
S.~Kittipiyakul and T.~Javidi, ``Relay scheduling and cooperative diversity for
  delay--sensitive and bursty traffic,'' in \emph{45th Annual Allerton
  Conference on Communication, Control, and Computing}, Monticello, Illinois,
  USA, Sep. 2007.

\bibitem{SenErkAaz1}
A.~Sendonaris, E.~Erkip, and B.~Aazhang, ``User cooperation diversity. part i.
  system description,'' \emph{{IEEE} Trans. Commun.}, vol.~51, no.~11, pp.
  1927--1938, 2003.

\bibitem{GamCaiDamARQ}
H.~El~Gamal, G.~Caire, and M.~O. Damen, ``The {MIMO} {ARQ} channel:
  diversity--multiplexing-delay tradeoff,'' \emph{{IEEE} Trans. Inf. Theory},
  vol.~52, no.~8, pp. 3601--3621, Aug. 2006.

\end{thebibliography}
\end{document}